\documentclass[oldversion]{aa} 
\usepackage{natbib}
\usepackage{graphicx}
\usepackage{amsmath}
\usepackage{txfonts}
\usepackage{footmisc}

\begin{document}
   \title{The X-ray puzzle of the L1551~IRS\,5 jet}
   \authorrunning{P. C. Schneider et al:}
   \titlerunning{X-rays from the L1551~IRS\,5 jet}

   \author{P. C. Schneider$^1$
          \and
          H. M. G\"unther$^2$
          \and
          J. H. M. M. Schmitt$^1$
          }

   \institute{$^1$\,Hamburger Sternwarte,
              Gojenbergsweg 112, 21029 Hamburg, Germany \,
              \email{cschneider/jschmitt@hs.uni-hamburg.de}\\
              $^2$\,Harvard-Smithsonian Center for Astrophysics, 60 Garden St., Cambridge, MA, USA  \email{hguenther@head.cfa.harvard.edu}
             }

   \date{Received .. / accepted ..}

  \abstract
   {
   Protostars are actively accreting matter and they drive spectacular, dynamic outflows, which evolve on timescales of years. X-ray emission from these jets has been detected only in a few cases and little is known about its time evolution. We present a new \textit{Chandra} observation of L1551~IRS\,5's jet in the context of all available X-ray data of this object. Specifically, we perform a  spatially resolved spectral analysis of the X-ray emission and find that (a) the total X-ray luminosity is constant over almost one decade, (b) the majority of the X-rays appear to be always located close to the driving source, (c) there is a clear trend in the photon energy as a function of the distance to the driving source indicating that the plasma is cooler at larger distances and (d) the X-ray emission is located in a small volume which is unresolved perpendicular to the jet axis by \textit{Chandra}.
   
   A comparison of our X-ray data of the L1551~IRS\,5 jet both with models as well as X-ray observations of other protostellar jets shows that a base/standing shock is a likely and plausible explanation for the apparent
   constancy of the observed X-ray emission.  Internal shocks are also consistent with the observed morphology if the supply of jet material by the ejection of new blobs is sufficiently constant.
   We conclude that the study of the X-ray emission of protostellar jet sources allows us to diagnose the innermost regions close to the acceleration region of the outflows.

   }

   \maketitle
%

\section{Introduction}
During the early stages of star formation the protostar is deeply embedded and therefore usually invisible at optical wavelengths; only infrared and radio emission, and potentially hard X-rays, penetrate the dense circumstellar environment. 
Yet, outflows escape the protostellar envelope and are detectable at various energy bands, thereby announcing the birth of a new star. 
At later stages of stellar evolution, when the accretion proceeds from a disk and the star becomes visible in the optical, outflow activity is still observed. However, the driving mechanism of these outflows remains elusive; neither the acceleration nor the collimation of the outflow are currently fully understood, magneto-centrifugally launched disk winds, with a possible stellar contribution are currently debated \citep[e.g.][]{Ferreira_2006}.

The most spectacular manifestations of these outflows are the condensations/shocked-regions termed Herbig-Haro (HH) objects. The proper motion of these knots within the outflows is generally in the range of a few 100$\,$km$\,$s$^{-1}$. Protostellar jets are intrinsically dynamic objects; their evolution is observable on timescales of a few years and models with variable ejection velocities can successfully explain some features of these jets \citep[e.g.][]{Raga_2010}. In these models, the overtaking of small slow blobs by faster more recently emitted blobs leads to shock fronts with shock velocities on the order of the amplitude of the velocity difference.

HH objects with X-ray emission are a recently discovered phenomenon. About ten such objects have so far been detected among the hundreds of known HH objects; shock velocities around 500~km\,s$^{-1}$ are required to heat material to X-ray emitting temperatures in terms of simple shock models. \object{HH~2} \citep{Pravdo_2001N} and the jet of \object{L1551~IRS\,5}  \citep[\object{HH~154},][]{Favata_2002,Bally_2003} marked the starting point of the X-ray discoveries. These outflows are driven by deeply embedded protostars (or their accretion disks). However, X-rays from the outflows are observed also from more evolved objects. The single classical T~Tauri star \object{DG Tau} shows a complex X-ray morphology: There is the outer X-ray jet emission complex, the inner X-ray emission region located at a few 10~AU from the star and the stellar X-ray emission \citep{Guedel_2008, Schneider_2008, Guenther_2009}.

In order to achieve high shock velocities within the jets, strongly varying outflow velocities are required to reach X-ray emitting temperatures.
The models by \citet{Bonito_2010a} require ejection velocities of a few 1000~km\,s$^{-1}$ in order to be reconciled with the X-ray observations. Such high velocities have not been detected in UV, optical or IR observations. As observable knots must experience at least one internal shock, these models predicted proper motion of the emission region with only a fraction of the initial flow velocity and are therefore also compatible with the observations in the UV, optical and IR. Also, only a fraction of the total mass-loss carried by the outflow is probably required to explain the observed X-rays. Therefore, part of the high velocity material might have escaped detection. 
It is not clear, if the low number of detections is caused by the low X-ray surface brightness of these jets or if only a few outflows actually emit X-rays at all.

With our third epoch high-resolution $Chandra$ observation of the L1551~IRS\,5 jet (HH~154) we aim to determine the time-variability of the X-ray emission in order to constrain the origin of the X-ray emission. Our article is structured as follows: We introduce in the L1551~IRS\,5 region in the next section.
In sect.~\ref{sect:dataAna} the observations and the data analysis are described. We then proceed to our results (sect.~\ref{sect:results}), which are discussed in sect.~\ref{sect:disc}. We review the implications of these results on current models in sect.~\ref{sect:models} and close with a summary in sect.~\ref{sect:concl}. 

As proposed by \citet{Bonito_2010a} throughout the text  the term ``knot'' describes a region of enhanced emission while ``blob'' refers to a moving gas clump which is not yet shocked, i.e. not observable in X-rays.

\section{L1551~IRS\,5: Overview of previous observations}
\begin{figure}[t!]
  \centering
   \fbox{\includegraphics[width=0.4\textwidth]{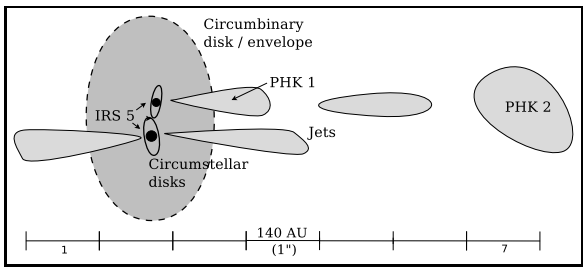}}
   \caption{Sketch of the region around the L1551 IRS~5 sources showing only a few components of the system (scales are only approximately preserved). The light gray regions indicate the jets. The inner two [Fe {\sc ii}] emission knots (PHK1, PHK2) from \citet{Pyo_2009} are marked, the circumbinary disk is shown in dark gray. Note that the northern optical jet might actually be driven by the southern binary component (only one counter-jet is shown). The one arcsec wide stripes are used later in the article (sect.~\ref{sect:morph}) and are shown here for reference. \label{fig:overview0} }
\end{figure}
The Lynds~1551 (L1551) star forming region \citep{Lynds_1962} is located at the southern end of the Taurus region at a distance  of approximately 140 pc; within this region, a number of protostellar objects and associated outflows have been found \citep[e.g.][]{Hayashi_2009}. Fig.~\ref{fig:overview0} shows a sketch of the immediate region around IRS\,5.
\subsection{The L1551~IRS~5 sources}
The term HH~154 describes the jet emanating from the sources collectively called L1551~IRS~5 \citep{Strom_1976}. The infrared source IRS\,5 consists at least of a protostellar binary system and additional components are possible \citep[see, e.g., the VLA data of ][]{Rodriguez_1998}. 
Each core is surrounded by its own circumstellar disk ($\sim 10\,\mbox{AU}$) and the complex is again embedded within a large envelope \citep[$\sim10000\,\mbox{AU}$;][]{Fridlund_2002}.
All central sources are hidden by substantial absorption of at least $A_V>20\,\mbox{mag}$, but probably as much as 150~mag \citep[e.g.][]{Snell_1985,Campbell_1988}. Therefore, the masses of the protostars are uncertain, but from spectral mapping of the reflected light escaping the envelope \citet{Liseau_2005} derived masses of 0.3 and $0.8\,M_\odot$.  These masses are consistent with the total system mass of 1.2\,$M_\odot$ derived by \citet{Rodriguez_2003}.
The separation of the two sources, and consequently their jets, is only 0.35\arcsec~(50~AU). We will use the term IRS~5 for both sources jointly although most of the time the source of the northern jet is considered.

\subsection{The L1551~IRS~5 jet (HH\,154)}
A double-lobed CO structure around L1551~IRS~5 was first detected by \citet{Snell_1980}. Subsequent high resolution optical and near-infrared observations revealed two separate, westwards directed jets emanating from the immediate region around the two VLA sources. They can be traced out to about 3\arcsec~\citep[420~AU, e.g.][]{Fridlund_1998, Itoh_2000}, where they become indistinguishable \citep{Fridlund_1998,Pyo_2003}. The inner most part is only observable at radio wavelengths \citep[e.g.][]{Rodriguez_2003}. Since the northern jet is brighter and faster, it is believed to be responsible for the Herbig-Haro (HH) objects further downstream at distances between a few and 12\arcsec~(1700~AU). The jet inclination has been estimated to $\sim 45^\circ$ \citep{Fridlund_2005}.

At a distance of about 3\,arcsec away from IRS~5 proper motion measurements of individual knots have been carried out on a baseline of 30~years, revealing substantial motion of individual knots; the inner knots show the highest projected space-velocities of up to 300~km/s \citep{Fridlund_2005, Bonito_2008}. These values are approximately consistent with highest (projected) blue shifted emission of up to 430~km/s \citep{Fridlund_2005}. 
However, high-resolution near-infrared [Fe II]~1.644~$\mu$m observations of the inner part showed an emission complex which is virtually constant over four years \citep{Pyo_2003,Pyo_2005, Pyo_2009}. The position-velocity diagrams (PVDs) show that the low-velocity component (LVC, $v<200\,$km$\,$s$^{-1}$) dominates the emission out to almost two arcsec, where the high-velocity component (HVC, $200\,\mbox{km$\,$s$^{-1}$} <v\lesssim\,450$km/s)  becomes dominant.
High-resolution Hubble Space Telescope images taken in small band-pass filters (e.g., H$\alpha$, [S~{\sc ii}]) can be explained by a light jet (i.e., less dense than the ambient medium), hitting into a denser ambient medium \citep{Fridlund_1998,Hartigan_2000}. Spectroscopically, the outer knots show a line-width of 110~km$\,$s$^{-1}$, densities from a few $10^3\,$cm$^{-3}$  to $8\times10^3\,$cm$^{-3}$ and an excitation  rising with decreasing distance to IRS~5 \citep[][and references therein]{Liseau_2005}.
Concerning the nomenclature, the visual knots are designated F, E and D \citep[in increasing distance from IRS~5, see][]{Fridlund_2005}, the inner near-infrared knots are termed PHK~1\dots3 \citep{Pyo_2003} in increasing distance to IRS~5 (cf. Fig.~\ref{fig:overview0}). Knot~D coincides with PHK~3 and knot~F with PHK~2.
Whether IRS~5 is also driving HH~28 and HH~29, which are located further downstream of HH~154, is not yet clear \citep{Devine_1999}, possibly L1551~NE is their driving source.

Throughout the text we will use the term HH\,154 for all outflow parts associated with the L1551~IRS\,5 jet.

\subsection{X-rays from HH~154}
\begin{figure*}[t!]
  \centering
   \fbox{\includegraphics[width=0.3\textwidth]{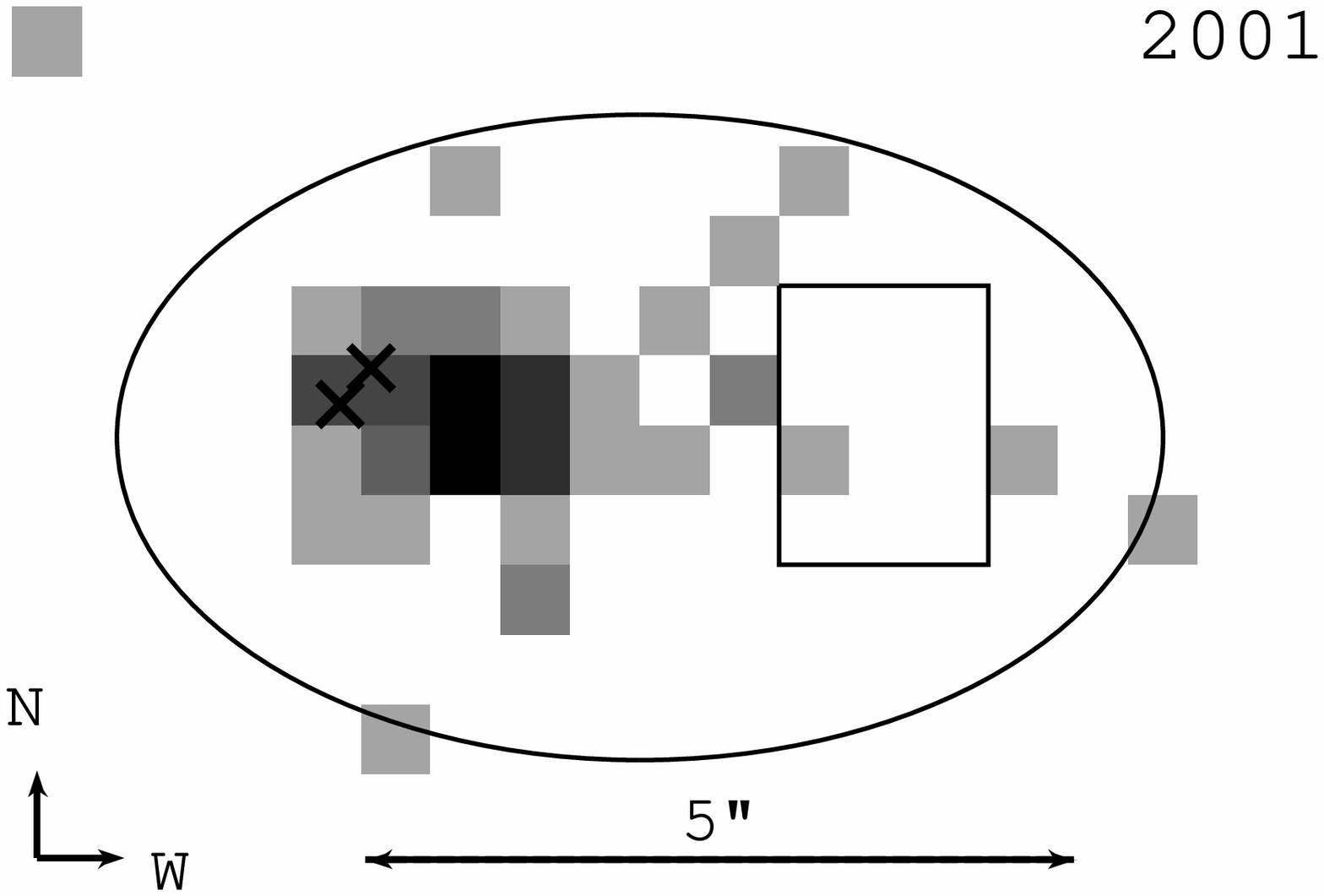}}
   \fbox{\includegraphics[width=0.3\textwidth]{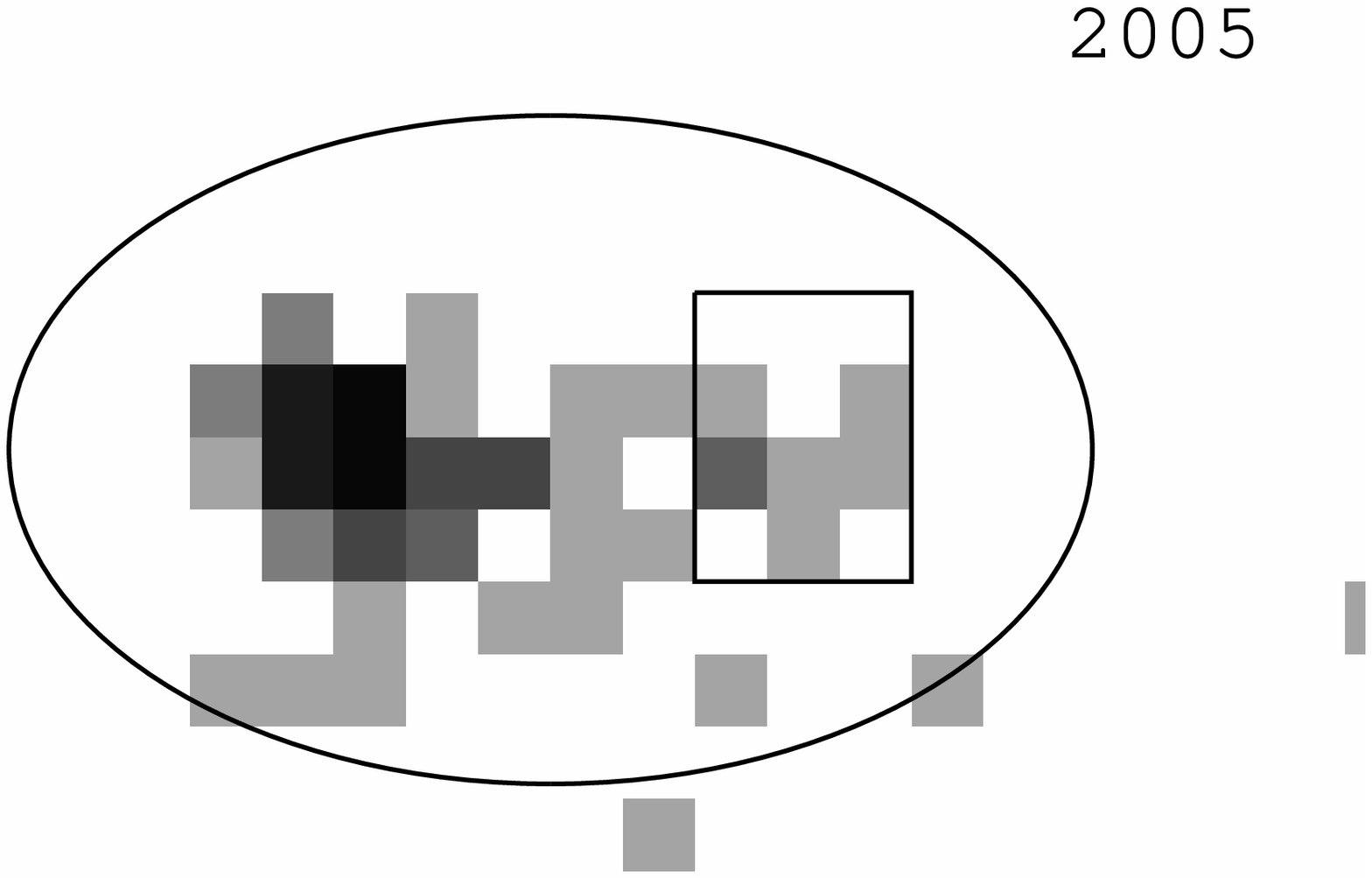}}
   \fbox{\includegraphics[width=0.3\textwidth]{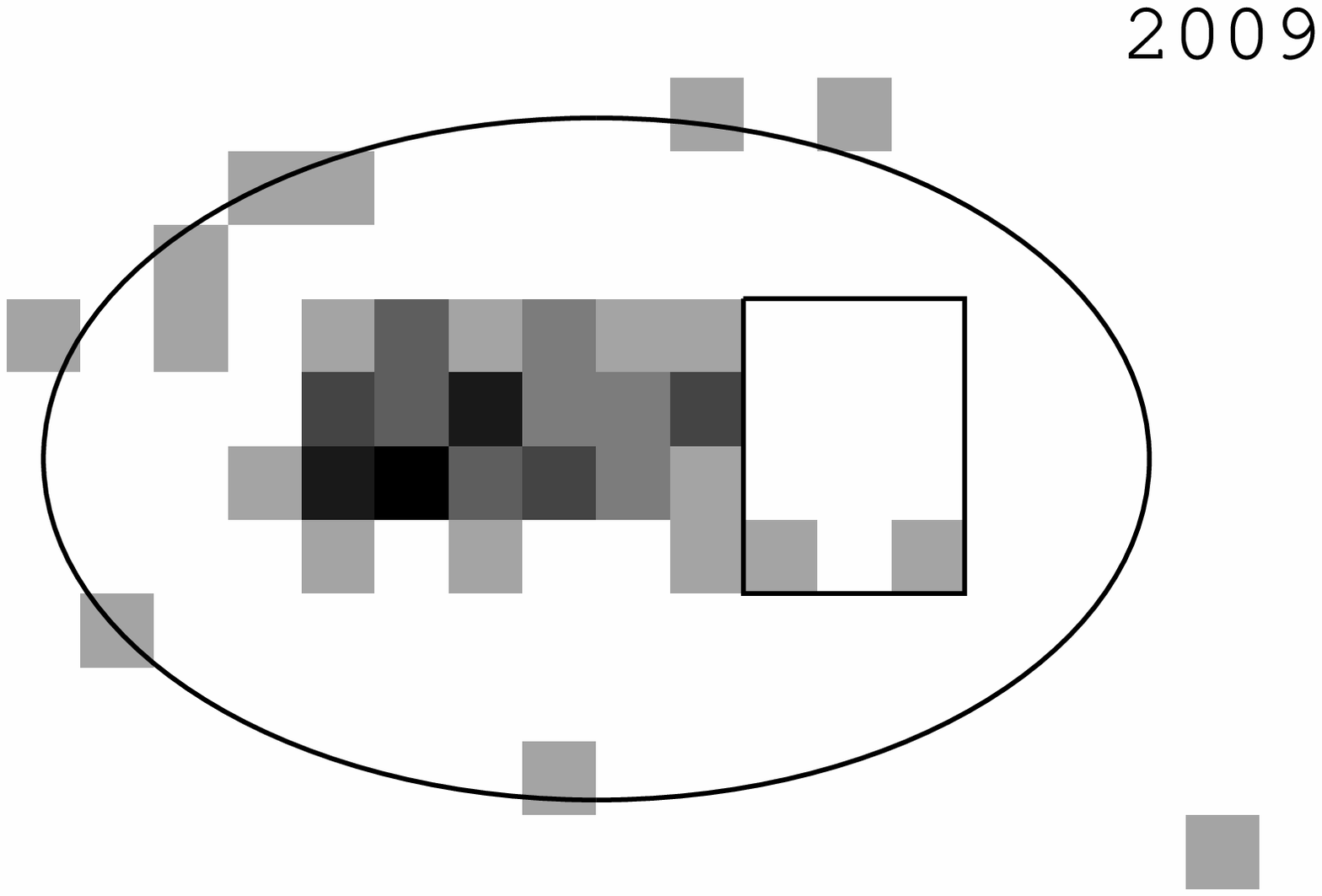}}
   \caption{The ellipse for the global plasma properties, also, the location of the 2005 ``knot'' is shown. The crosses indicate the position of the radio sources and their size approximately indicates the position uncertainty. \label{fig:overview} }
\end{figure*}

HH~154 was first discovered as an X-ray source  by \citet{Favata_2002} from an observation with $XMM$-Newton. Despite of the large PSF of $XMM$-Newton ($\sim15\;$\arcsec), the authors correctly concluded from the spectral properties of the photons that the X-ray emission cannot be associated with IRS~5 and proposed knot~D as the source of the X-ray emission; knot~D is the brightest optical knot and probably the current terminal working surface. Higher resolution $Chandra$ observations (see Fig.~\ref{fig:overview}) revealed that the X-ray source is actually located further inward towards IRS~5 at a distance of 0.5-1.0\arcsec~ from IRS\,5 \citep{Bally_2003}.
A second epoch $Chandra$ exposure showed a somewhat different morphology of the X-ray emission being more elongated than the 2001~ACIS data. This elongation has been interpreted as a moving X-ray knot with a projected space velocity of about 330~km$\,$s$^{-1}$ \citep{Favata_2006}.

Based on their initial detection of X-rays from HH~2 \citet{Pravdo_2001N} proposed shocked, high-velocity knots as the  explanation of the observed X-rays. The first analytical description of this process was presented by \citet{Raga_2002} and \citet{Bonito_2004} performed the first numerical hydrodynamic models with an emphasis on the X-ray emission. These models have been extended towards variable blob ejection velocities by \citet{Bonito_2010b, Bonito_2010a} and their analysis revealed that very fast blob velocities of more than 1000~km/s are needed in order to explain the observations by shock heating, i.e., by internal shocks occurring when fast blobs overtake slower ones. An ejection ``period'' of two years matched the X-ray observations best.

\section{Observations and data analysis\label{sect:dataAna}}
Table~\ref{tab:Observations} lists all the available X-ray observations of HH~154.
We used the ACIS-S detector for the third epoch $Chandra$ exposure, since the back-illuminated ACIS-S chip has a higher sensitivity at lower energies than the front-illuminated ACIS-I CCDs. With the VFAINT mode, this setup provides a similar sensitivity as the longest ACIS-I exposure for the energies at hand;  and an even higher sensitivity for plasma at cooler temperatures as expected for individual knots moving outwards and cooling.

We used \emph{CIAO} version 4.2 throughout the data analysis and followed the science threads on the \emph{CIAO} webpage\footnote{\texttt{http://cxc.harvard.edu/ciao/}}.
The \mbox{ACIS-S} observation was reprocessed to account for the VFAINT-mode\footnote{\texttt{http://cxc.harvard.edu/ciao/threads/createL2/}}. We experimented with pixel randomization, but since the relevant scales are usually at least twice the detector pixel scale, the effect of pixel randomization is virtually negligible. Therefore, we used the standard processing including pixel randomization. We explicitly note in the text where we expect this assumption to be invalid.

\begin{table}
\begin{minipage}[h]{0.49\textwidth}
\caption{Analysed X-ray observations of HH~154.}\label{tab:Observations}
\centering
\setlength\tabcolsep{2pt}
\renewcommand\footnoterule{}
\begin{tabular}{l c c c c c}

\hline\hline
Date & Observatory & Setup & Obs-ID & exp. time\\
\hline
2000-09-09 & XMM-Newton & -- & 0109060301 & 56~ks\\
2001-07-23 & $Chandra$ & ACIS-I & 1866 & 80~ks\\
2004-03-\dots\footnote{This dataset consists of 11 short exposure distributed over six days in March 2004 where HH~154 is off-axis by about 2 arcmin.} & XMM-Newton & --  &  0200810201\dots & 107~ks \\
2005-10-27 & $Chandra$ & ACIS-I & 5381 & 98~ks\\
2009-12-29 & $Chandra$ & ACIS-S & 11016 & 66~ks\\
\hline
\end{tabular}
\end{minipage}
\end{table}

In order to improve the astrometric accuracy, the three $Chandra$ observations of HH~154 were aligned by calculating the centroids of the brightest sources detected by the \emph{CIAO} tool \texttt{celldetect} and the photon events were reprojected so that the mean offset between these centroids vanishes. 
We use the offset obtained from the three brightest sources weighted by the square-root of their count-number. These sources are also members of the Taurus star forming region and located on the ACIS-S part of the CCD array. Using more sources (detected with S/N$>$3) and equally weighted or photon number weighted means changes the offset by less than 0.5~pixel (0.25\arcsec). Thus, relative positions are at least accurate to within one pixel (0.49\arcsec). For a cross-check of the positions we calculated the centroid positions of HH~154 using the photons within a circle of 2.5 pixel radius centered on the brightest emission peak in the energy range 0.5 - 3.0 keV. They coincide to within approximately 0.3\arcsec, well within our estimated accuracy. Note that these centroids should coincide only in case of a stationary source. We give distances relative to the radio position with respect to the nominal positions of the 2001 observation where the comparison sources show a good agreement with their optical positions but note that this position is accurate only within 0.5\arcsec.

We also retrieved the archival {\it XMM}-Newton data (Obs-ID 0109060301) from 2000 for a spectral crosscheck and eleven exposures during March 2004 (Obs-ID 0200810201\dots0200811301), where HH~154 is located about two arcmin off-center for a luminosity check. SAS~9.0\footnote{\texttt{http://xmm.esa.int/sas/}} was used for the analysis of the XMM-Newton data. We extracted the source photons within a circle of 15\arcsec~ around the source position of HH~154 for the March 2004 exposures and derived the background from a nearby source-free region. We concentrated on the MOS data since the PN suffers high background levels (using standard filters only 40~ks PN on-time remain). However, both count-rates agree within their respective 1$\sigma$ ranges. We converted the count rate to  luminosity by assuming the same spectral properties as during the 2001 XMM-Newton observation.

\begin{figure}
  \centering
   \includegraphics[width=0.49\textwidth]{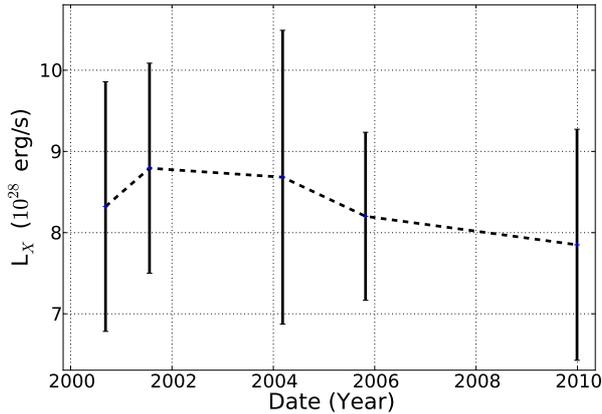}
   \caption{Unabsorbed X-ray luminosity (0.5-10\,keV) of HH~154 including the XMM-Newton data for the 2000 and 2004 data points. As the XMM-Newton data lacks sufficient spatial resolution to resolve the X-ray emission, the displayed luminosity pertains to the total observed X-ray emission for all datasets.\label{fig:lc} }
\end{figure}

\section{Results \label{sect:results}}
The X-ray images of HH~154 in the 0.5--3.0~keV energy band for the three available $Chandra$ observations are shown in Fig.~\ref{fig:overview}. To extract photons we use an ellipse with semi-axis lengths of 3.7\arcsec{} and 2.3\arcsec, respectively, whose semi-major axis is aligned with the centroid in declination, i.e., which is aligned approximately with the jet axis and  contains all photons attributable to the X-ray emission of HH~154. Using a nearby source free background region (no X-ray sources detected nor a 2MASS \citep{2MASS} or SIMBAD\footnote{\texttt{http://simbad.u-strasbg.fr/simbad/}} source known), the expected background ($E_{photon}=0.5\dots3.0\;$keV) within the ellipse is 0.7 photons for the 2001 observation, 1.0 for the 2005 observation and 1.4 in the 2009 observation. 

\subsection{Energetics}
We used XSPEC\,v12.5.0 \citep{XSPEC} for the spectral modeling and assumed that the observed material can be described by optically thin thermal plasma emission \citep[APEC, ][]{APEC} and included absorption by neutral gas along the line of sight in our fits. We set the abundance to half the solar value \citep[using][]{grsa} since the XEST survey of the Taurus region found on average a sub-solar metallicity \citep{XEST}.

Figure~\ref{fig:lc} shows the unabsorbed X-ray luminosity of HH~154 during the last decade. It clearly indicates that the luminosity in the HH~154 region appears constant and deviations from the mean value do not exceed 22\%.

\begin{figure}
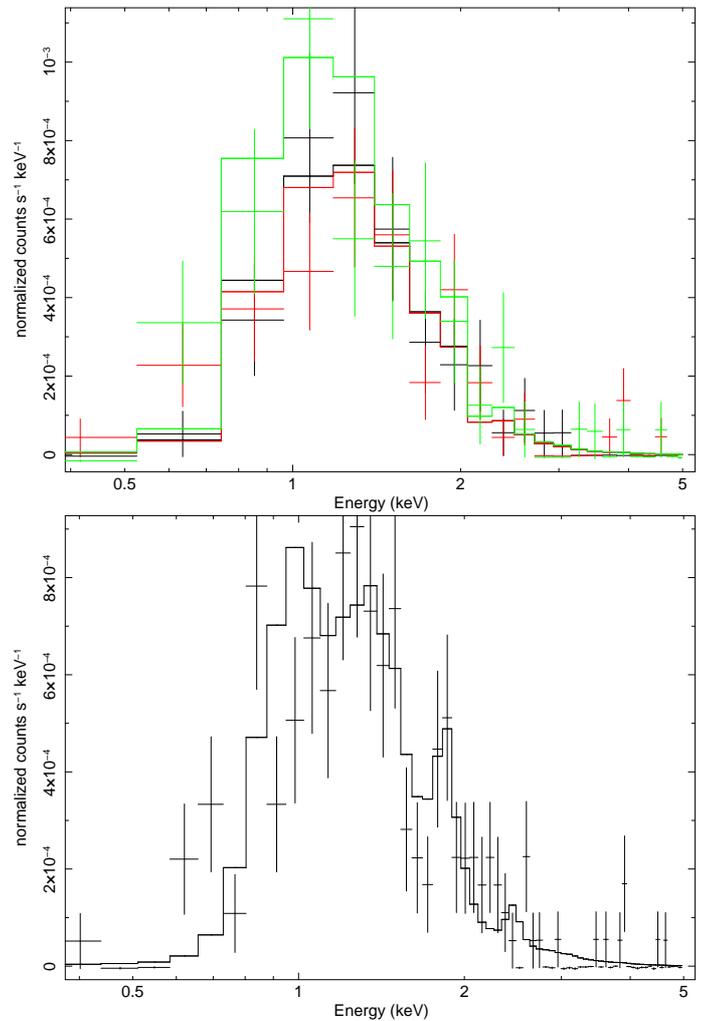

  \centering
   \includegraphics[height=0.49\textwidth, angle=270]{cs_fg4a.ps}
    \includegraphics[height=0.49\textwidth, angle=270]{cs_fg4b.ps}
   \caption{Spectrum of all photons within the ellipse (see Fig.~\ref{fig:overview}). {\textbf Top:} Individual spectra, {\textbf Bottom:} Co-added spectra (the Si-lines around 1.8\,keV are clearly visible).\label{fig:spectrum} }
\end{figure}

\subsubsection{Global plasma properties}
The spectra extracted using the photons within the ellipse are shown in Fig.~\ref{fig:spectrum}, however, we caution that the assumption of homogeneous plasma properties throughout the emitting region is probably not valid (see sect.~\ref{sect:localPlasma}).
Averaging over different plasma properties leads to an unstable fit with two solutions describing the data reasonably well; the two  possibilities are listed in Tab.~\ref{tab:XSPECresults1} (top). One solution is only weakly absorbed and requires rather high plasma temperatures. Although this solution is statistically favored, we regard this solution as physically less plausible due to the following reasons.
The Si lines at $\sim$1.9~keV are not reproduced by this model but clearly present in the data (bottom panel of Fig.~\ref{fig:spectrum}) and the low $N_H$ value contrasts the high absorption derived for the region close to the driving sources ($N_H \gtrsim 10^{23}\;$cm$^{-2}$) and along the jet axis \citep[$N_H \gtrsim 8\times 10^{21}\;$cm$^{-2}$, e.g.][]{Itoh_2000, Fridlund_2005}. In the following, we will therefore concentrate on the fit solution I with higher absorption and lower plasma temperature.
The plasma properties within the ellipse are compatible with each other for the individual exposures  (1$\sigma$). 
The discrepancy between the values given by \citet{Favata_2002} and our values results partly from the change of metallicity (the respective $1\sigma$ ranges overlap for solar metallicity).

In order to estimate an upper limit for the presence of cooler plasma, we added a second temperature component to the fit and fixed its temperature to 0.3\,keV (0.2\,keV). The 1$\sigma$ upper limit on the luminosity of this cool component is $1.4\times10^{29}$\,erg\,s$^{-1}$ ($3.0\times10^{29}$\,erg\,s$^{-1}$) when the absorption is forced to the value of the one temperature fit which also agrees with the optical/near-infrared value. Allowing the absorption to vary, the luminosity of the low temperature component decreases as the absorption also decreases for this two temperature component fit.

\subsubsection{Local plasma properties \label{sect:localPlasma}}
For a quantitative comparison of the three $Chandra$ observations we divide the region around HH~154 into $\sim1\arcsec$ (2 ACIS pixels) wide spatial bins as indicated in Fig.~\ref{fig:stripes}. For a point source about 80\% of the photons are located in these 1\arcsec{} wide stripes. This procedure is essentially a projection of the photon number onto the flow-axes (i.e., the x-axis). Figure~\ref{fig:stripes} shows the result for the individual exposures. Naturally, the exact values depend on the stripes used, therefore, we checked the results by shifting the stripes or using a different width of the stripes (about 90\% of the photons of a point-source would be included in a three pixel wide stripe). Any property which depends crucially on the choice of the stripes is regarded as an unphysical artifact (Tab.~\ref{tab:StripeProps}).

\begin{figure}
   \includegraphics[width=0.49\textwidth]{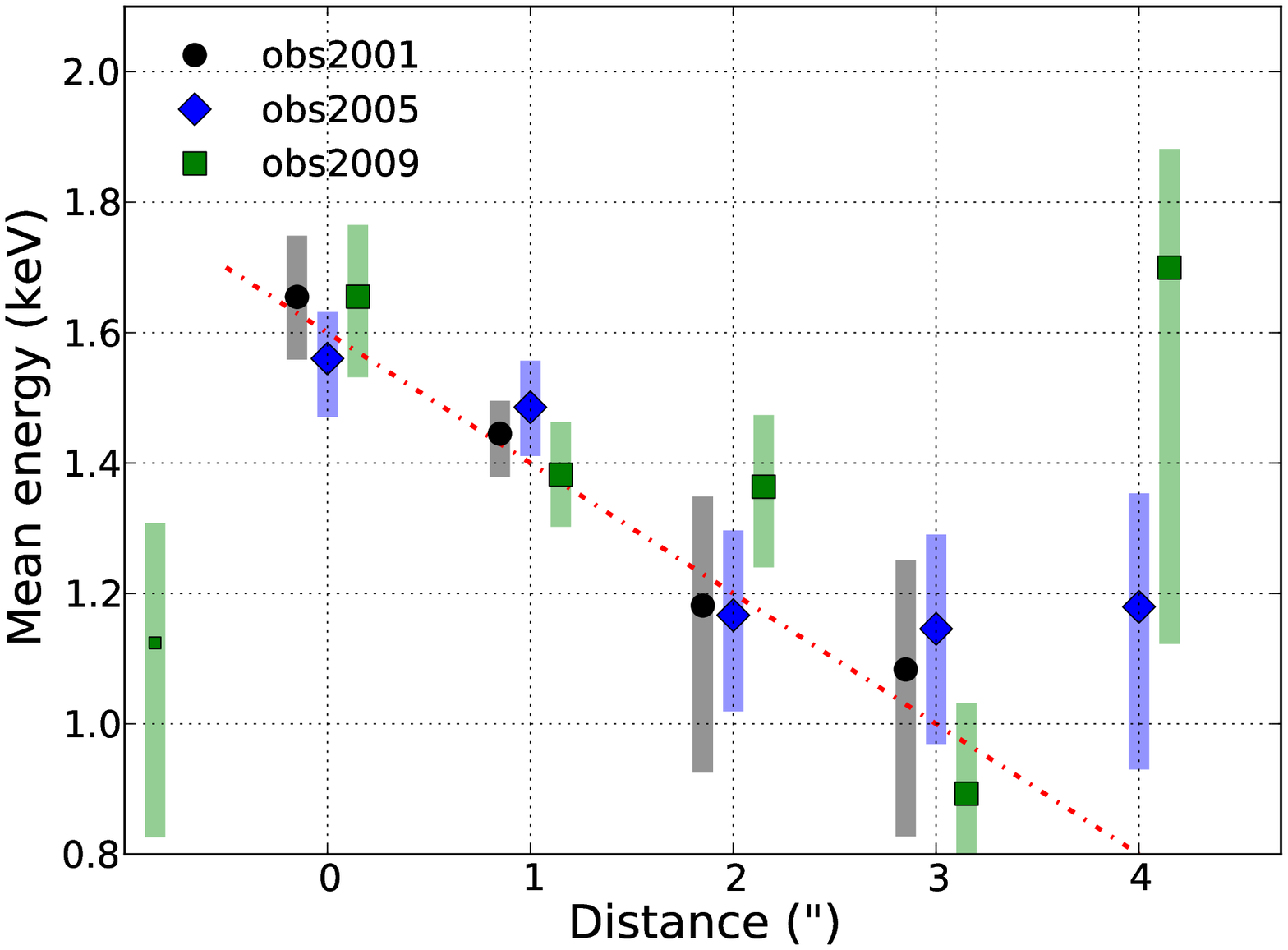}
   \includegraphics[width=0.49\textwidth]{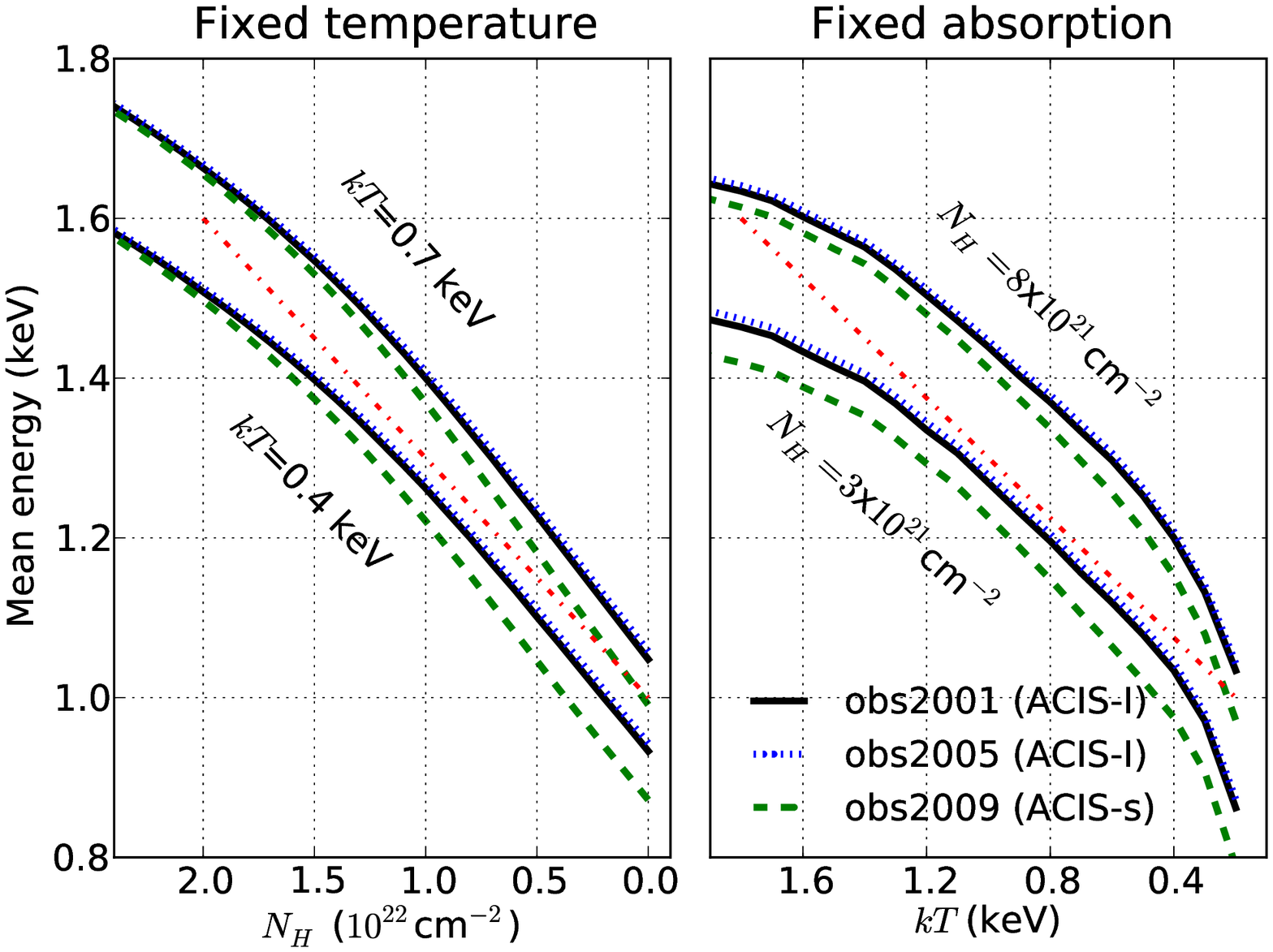}
   \caption{Trend of mean energies, stripes are those of Fig.~\ref{fig:stripes}. Photons used within 0.5-3.0 keV. The red dash-dotted line is intended to guide the eye in the lower panel. \textbf{Top:} Thick symbols indicate data points with more than three photons while the smaller symbols refer to data points with fewer photons. Errors are obtained using the simulations of sect.~\ref{sect:localPlasma}. \textbf{Bottom:} Simulations for fixed temperature and fixed absorption along the jet axis. \label{fig:meanE} }
\end{figure}

The mean energy of the photons in the individual strips along the flow axis is displayed in Fig.~\ref{fig:meanE} (top panel), showing a clear decrease of photon energy with increasing distance to the driving source(s).
The slope of the mean energy depends on the detector's spectral response and on the stripes used.
However, the spectral softening with increasing distance is independent of the detector response since only relative changes within the same detector are compared.

For a cross check of the trend of the mean energy, we divided the emission region into an ``eastern'' and a ``western'' region so that the component close to the driving source(s) is associated with the ``eastern'' region and the outer part of the emission with the ``western'' region (these two regions essentially split stripes 2-7 into 2 separate regions); Table~\ref{tab:XSPECresults1} lists the associated fit results. Since the total number of counts in the ``western'' region is only 42 when summed over all ACIS exposures ($E_{\mbox{photon}}=0.3$--$5.0\;$keV), we checked the results by fixing either the temperature or the absorption to the values obtained for the left part of the emission, this results in a decrease of the $N_H$ value consistent with Fig.~\ref{fig:meanE} and the temperature for the right emission component, respectively.

\begin{table}
\begin{minipage}[h]{0.49\textwidth}
\renewcommand{\arraystretch}{1.5}
\caption{Plasma properties of HH~154 with 1$\sigma$ errors. Spectra are binned to $\sim44\;$eV wide channels, i.e. about three times oversampling the intrinsic energy resolution of the ACIS detectors, we used c-stat and the energy range 0.5-5.0 keV. Unabsorbed fluxes (0.5-10. keV) are given. }\label{tab:XSPECresults1}
 \centering
\setlength{\columnsep}{0.2mm}
\renewcommand\footnoterule{}

\begin{tabular}{l c c c c}

\hline
\hline
& $N_H$       & $kT$  &       $EM$           & $L_X$  \\[-1mm]
& (10$^{22}\;$cm$^{-2}$) & (keV) & ($10^{51}\;$cm$^{-3}$) & ($10^{27}\;$erg/s)\\
\hline
\multicolumn{5}{c}{Composite spectrum (ellipse)}\\
Solution I   & 1.1 $\pm$ 0.1 & 0.6 $\pm$ 0.1 & 7.9$\pm2.2$ & 82.7$_{-21.8}^{+20.0}$\\
\multicolumn{1}{r}{obs2001\footnote{$N_H$ and $kT$ fixed to values from  co-added  spectra\label{note:fixed}}}
             & \multicolumn{2}{c}{}  & $8.6_{-1.8}^{+2.6}$& 89.7$_{-24.4}^{+24.3}$\\
\multicolumn{1}{r}{obs2005 \footref{note:fixed} }
             &   && $7.7_{-3.4}^{+2.3}$&81.6$_{-21.8}^{+21.7}$\\
\multicolumn{1}{r}{obs2009 \footref{note:fixed} }
             &   && 7.4 $_{-3.4}^{+2.2}$          & 77.6$_{-21.0}^{+21.4}$\\
Solution II & 0.2 $\pm$ 0.1 & 1.8 $_{-0.1}^{+0.3}$ & 1.9$\pm0.2$ & 18.4$_{-1.9}^{+2.1}$\\ 
\hline
\multicolumn{5}{c}{Individual spectra (ellipse)}\\
XMM (2000)     & 1.0 $\pm 0.1$         & 0.5 $\pm 0.1$        & 8.4$_{-2.0}^{+3.6}$ & 89.7$^{+29.2}_{-20.2}$\\
obs2001        & 1.1 $^{+0.2}_{-0.3}$  & 0.6$^{+0.4}_{-0.1}$  & 8.9$_{-5.3}^{+6.4}$ & 97.2$_{-33.2}^{+37.1}$ \\
obs2005        & 1.1 $\pm$ 0.2         & 0.6$\pm$ 0.1         & 7.8$_{-2.7}^{+3.5}$ & 79.8$_{-25.5}^{+28.7}$\\
obs2009        & 1.0 $\pm$ 0.2         & 0.6 $_{-0.1}^{+0.2}$ & 7.0$_{-2.9}^{+3.5}$ & 73.4$_{-29.7}^{+32.7}$\\
\hline
\multicolumn{5}{c}{Composite spectra}\\
Eastern region & 1.4$^{+0.2}_{-0.3}$ &  0.7$\pm$0.2 & 6.6$_{-0.1}^{+2.7}$ & 78.1\\
Western region & 0.6$_{-0.2}^{+0.3}$ & 0.3$\pm0.2$ & 0.4$_{-0.4}^{+0.9}$ & 5.0\\
\hline
\end{tabular}

\end{minipage}
\end{table}

Figure~\ref{fig:meanE}~(bottom panel) shows the change required in either the plasma temperature or the absorption to explain the trend of the mean photon energy; fixing one parameter requires large changes of the other parameter ($N_H=2\times10^{22}\rightarrow0\;$cm$^{-2}$ or $kT=1.2\rightarrow0.2\;$keV). 

The errors shown in Fig.~\ref{fig:meanE}~(top) were obtained by simulating spectra containing a specified number of photons and then calculating the mean energy range which contains 68~\% of the trials. For photon numbers larger than $\sim$20, the error ($\pm0.1\;$keV) depends only very weakly on the number of photons; furthermore, the error depends only weakly on the assumed plasma properties for the spectra at hand.

\subsection{Morphology along the jet axis \label{sect:morph}}
\begin{figure*}
  \centering
   \fbox{\includegraphics[width=0.3\textwidth]{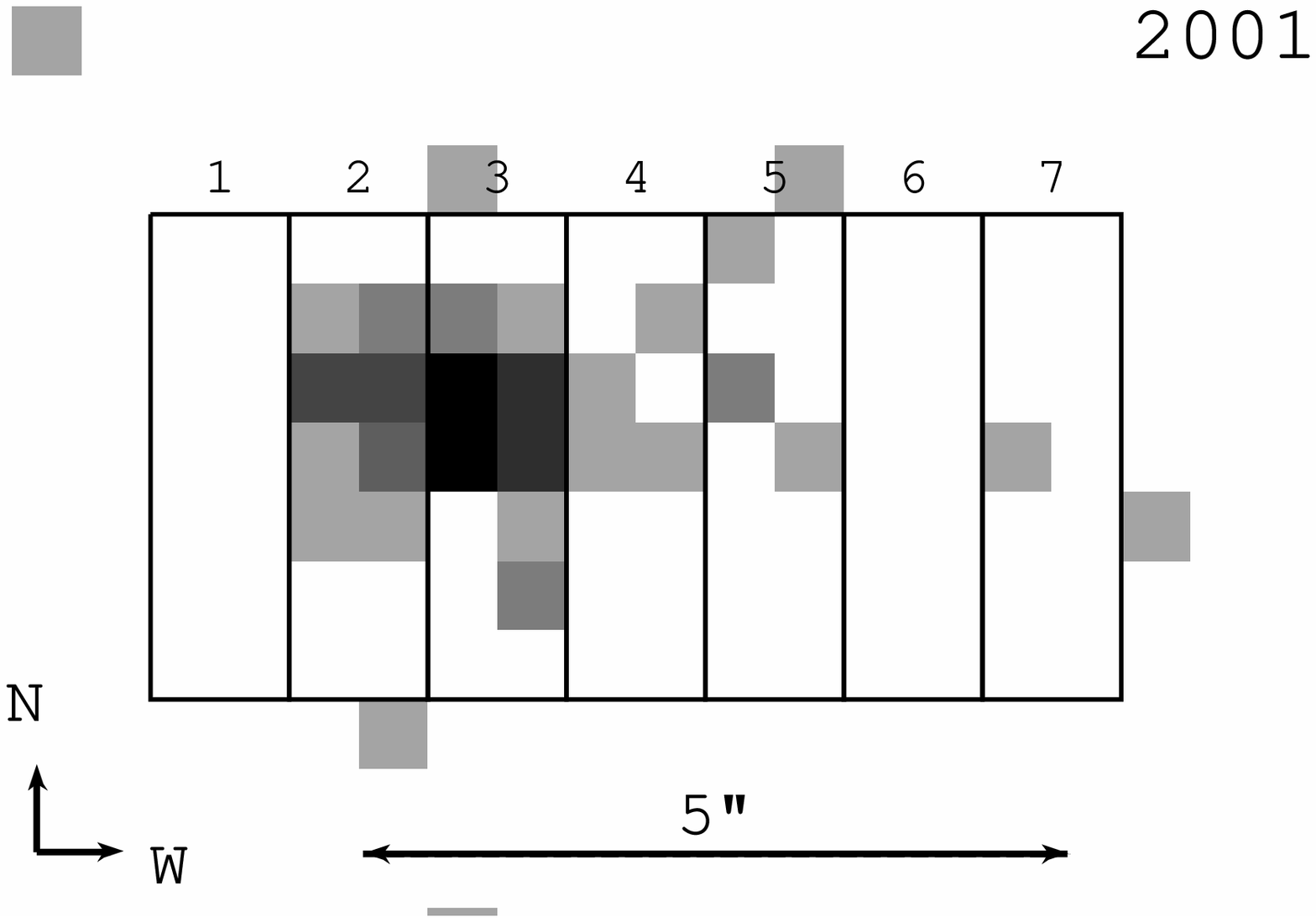}}
   \fbox{\includegraphics[width=0.3\textwidth]{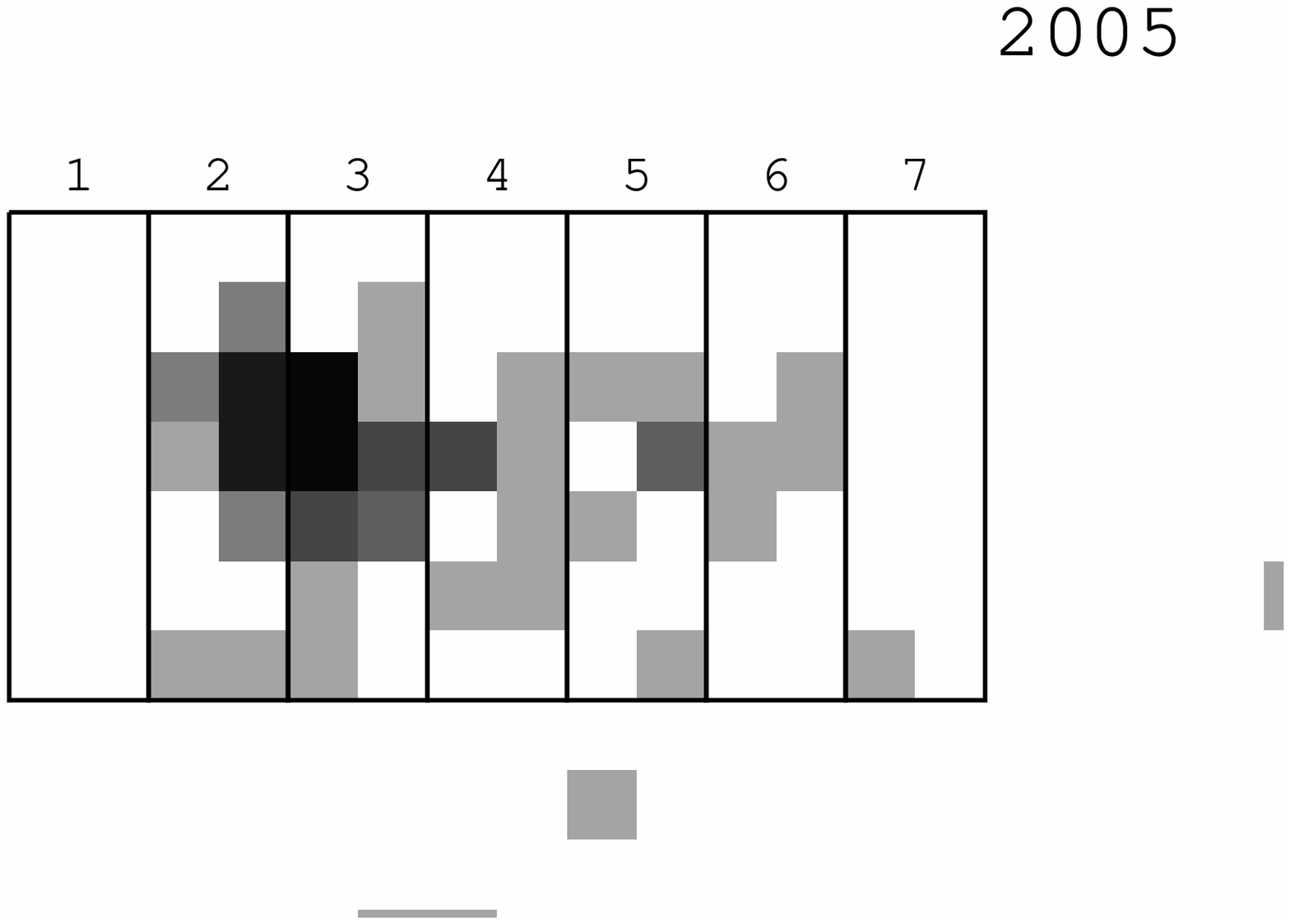}}
   \fbox{\includegraphics[width=0.3\textwidth]{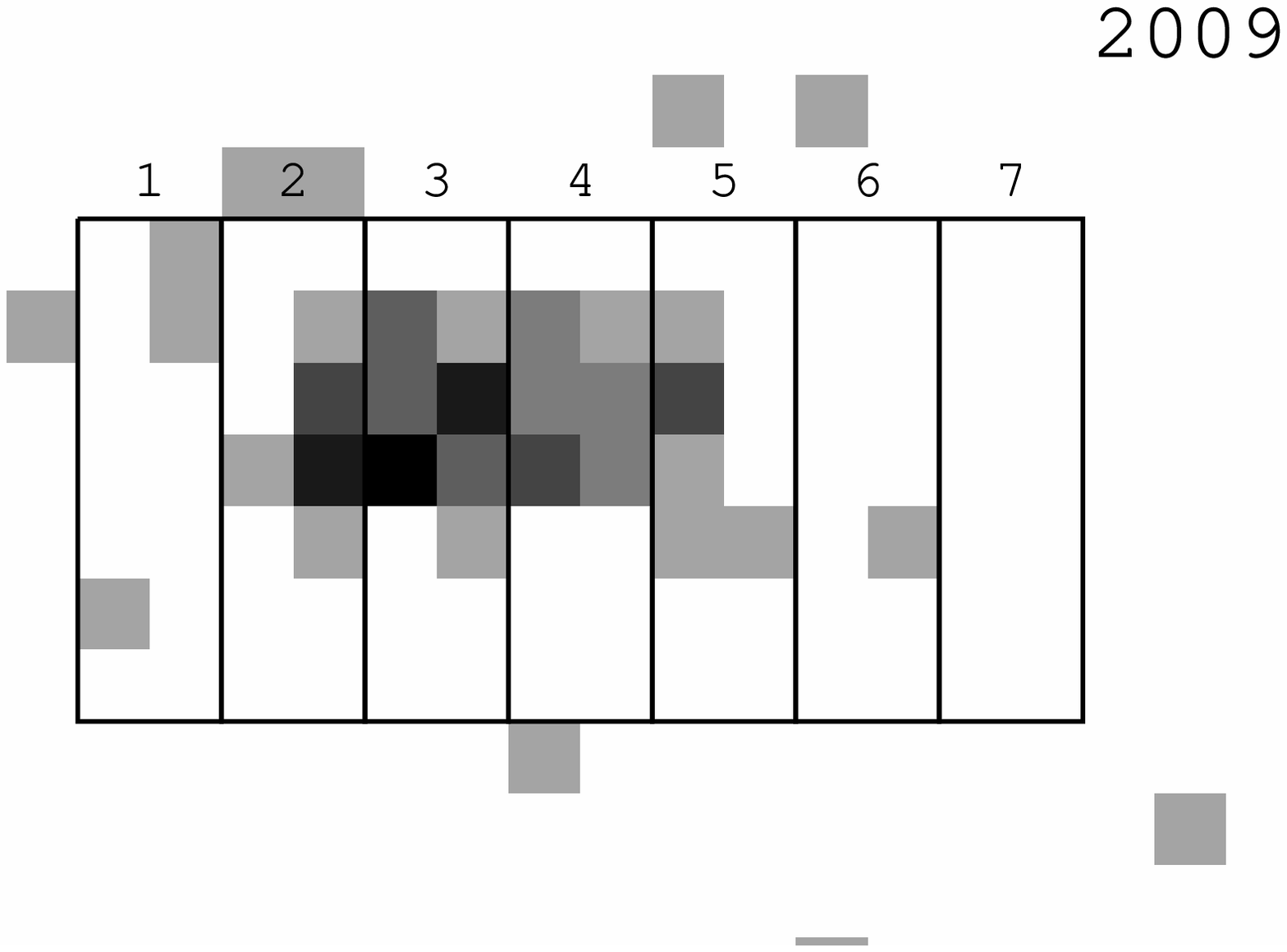}}
   \includegraphics[width=0.3\textwidth]{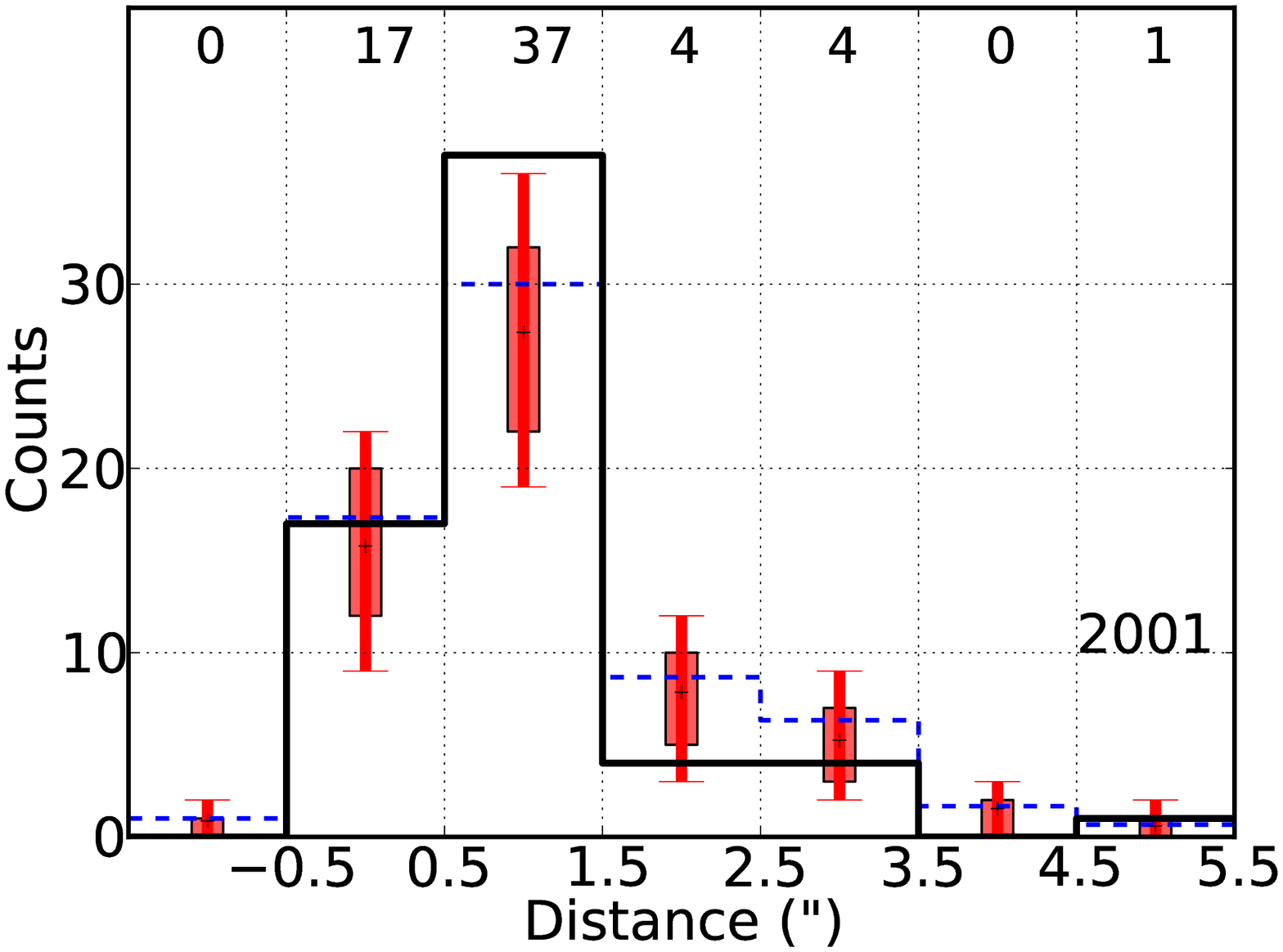}
   \hspace{3mm}
   \includegraphics[width=0.3\textwidth]{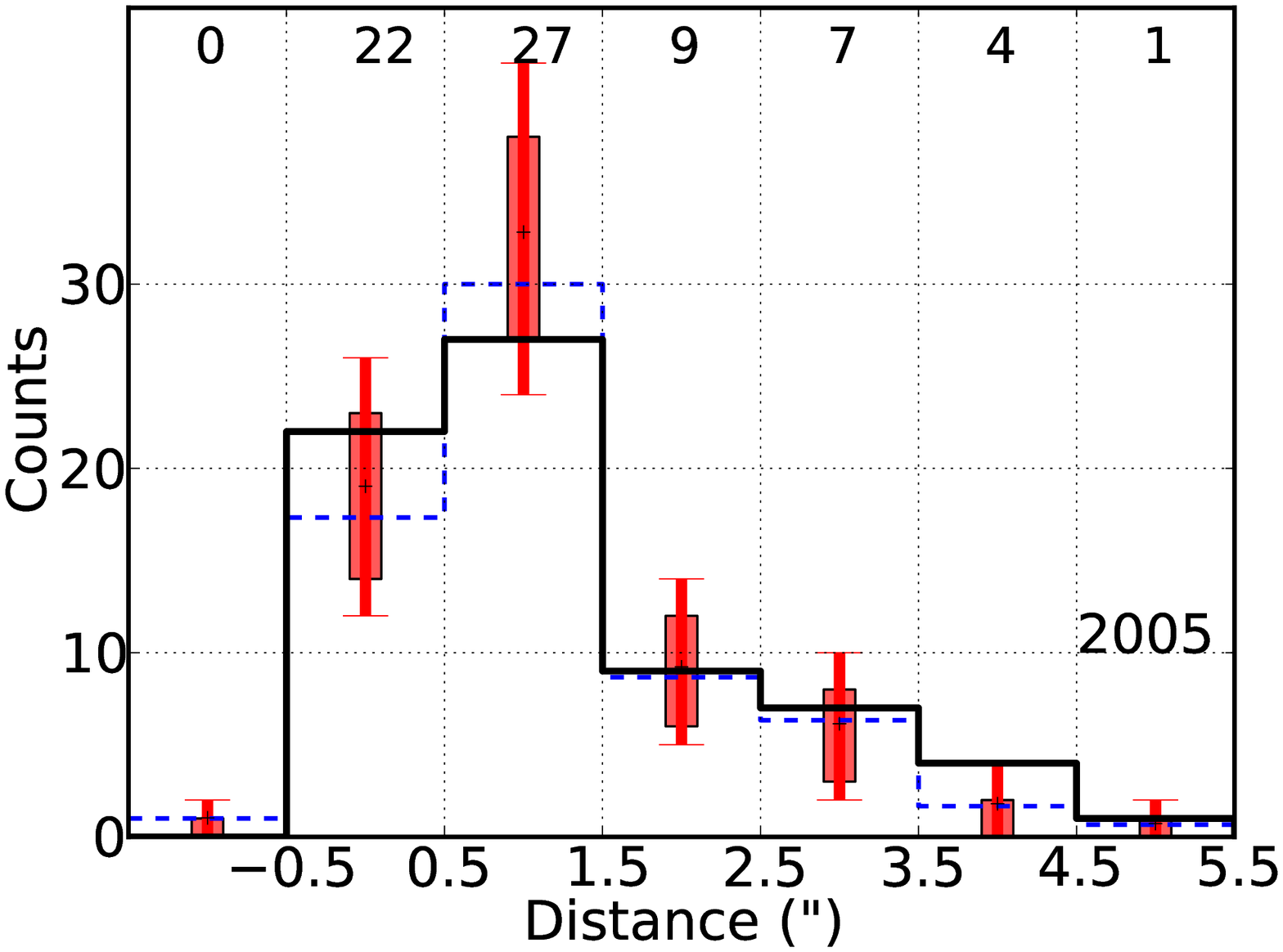}
   \hspace{3mm}
   \includegraphics[width=0.3\textwidth]{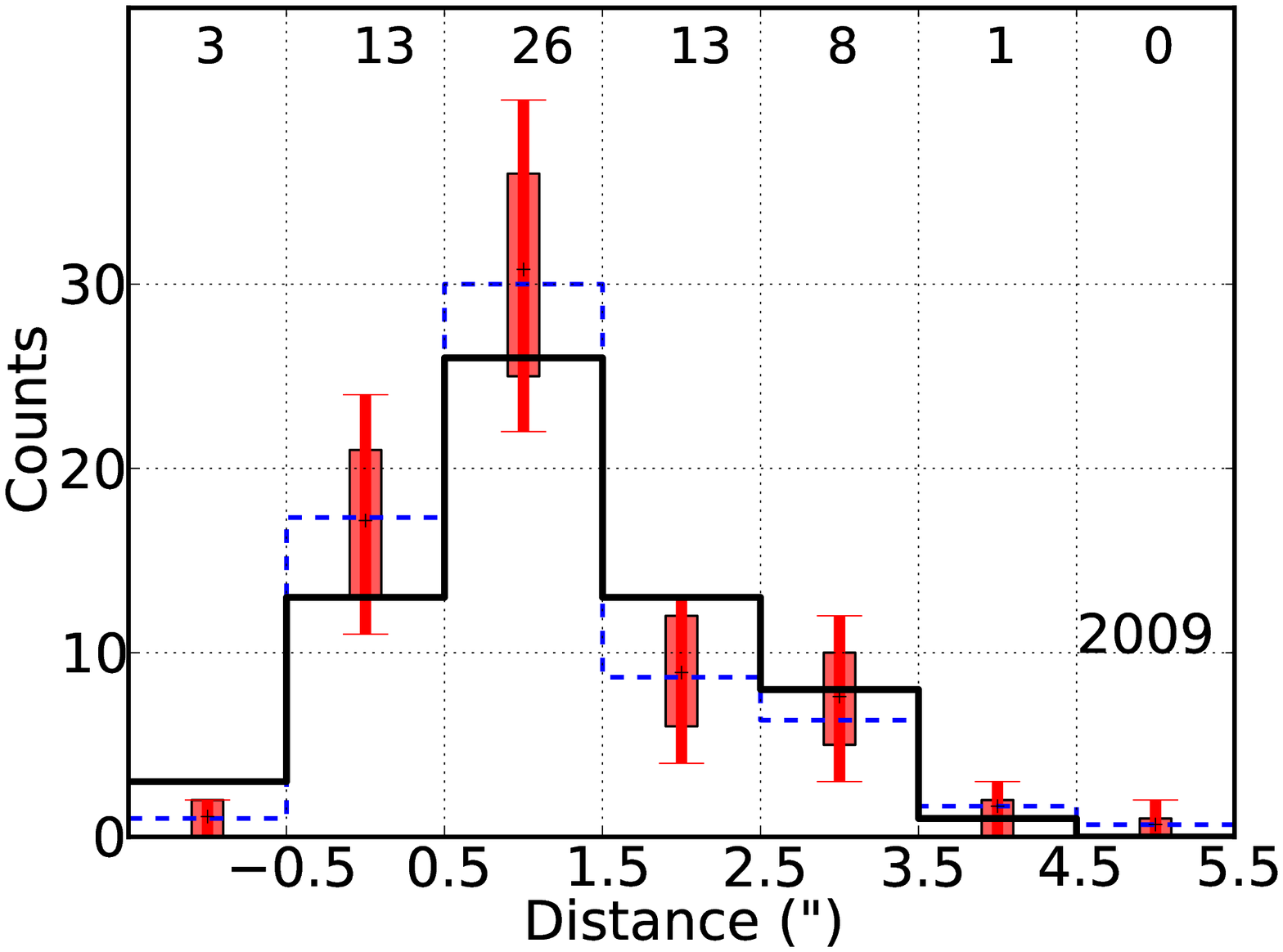}
   \caption{One example of stripes (1\arcsec width) used to extract count numbers, mean energies and spectra. In the bottom row, the projection onto the outflow direction (here assumed to coincide with the x-axis) is shown, the thick black line is the measured photon number, the thin blue dashed line indicates the mean photon number ignoring differences in the efficiency of the individual exposures. The red bars indicate the expectation value for a time constant emission, shown are the 70~\% and 90~\% probability ranges. \label{fig:stripes} }
\end{figure*}

The structure of the X-ray emitting region observed in 2009 neither resembles the structure present in 2001 or 2005; it is more extended than the 2001 structure, but does not show excess emission as far downstream as the 2005 exposure. The emission region close to the driving source, which is also present in all previous exposures, is most notable. The new ACIS-S observation does not show a clear knot westwards (downstream) of the main emission component as suggested by the 2005 image and there is no X-ray emission even further downstream as would be expected for a moving knot of constant luminosity. 
Note that the ACIS-S exposure is more sensitive to low energy photons than the 2005 ACIS-I exposure and that, according to Fig.~\ref{fig:stripes}, the photons soften with increasing distance to the driving source. Therefore, any emission with comparable properties as the photons attributed to the ``knot'' should be detectable with the 2009 ACIS-S observation.

\begin{table}
\caption{Stripe properties (mean energy is in keV).}\label{tab:StripeProps}
\centering
\setlength\tabcolsep{0pt}
\begin{tabular}{c l c c c c c c c c c c c c}
\hline\hline
Observation & & \multicolumn{12}{c}{Stripe}\\
& & 1 && 2 && 3 && 4 && 5 && 6\\
\hline
2001 & Counts      & 0   && 17  && 37  && 4   && 4   && 0\\
     & Mean energy & --  && 1.7 && 1.4 && 1.2 && 1.1 && -- \\
2005 & Counts      & 0   && 22  && 28  && 9   && 7   && 4 \\
     & Mean energy  & --  && 1.6 && 1.4 && 1.2 && 1.1 && 1.2\\
2009 & Counts      & 3   && 13  && 26  && 13  && 8   && 1\\
     & Mean energy  & 1.2 && 1.7 && 1.4 && 1.4 && 0.9 && 1.7\\
\hline
& & \multicolumn{12}{c}{Stripes shifted by 0.5\arcsec}\\
2001 & Counts      && 7   && 33  && 16  && 5   && 1   && 1\\
     & Mean energy && 1.7 && 1.5 && 1.3 && 1.2 && 1.2 && 1.0 \\
2005 & Counts      && 4   && 36  && 14  && 6   && 7   &&  3 \\
     & Mean energy  && 1.8 && 1.5 && 1.4 && 1.2 && 1.0 && 1.6\\
2009 & Counts      && 3   && 27  && 19  && 12  && 1   && 1\\
     & Mean energy  && 1.2 && 1.6 && 1.3 && 1.1 && 1.2 && 1.7\\
\hline
\end{tabular}
\end{table}

\begin{table}
\caption{Spectral models for sect.~\ref{sect:pNumbers}}\label{tab:models}
 \centering
\setlength{\columnsep}{0.2mm}
\begin{tabular}{l c c }
\hline
\hline
Model  &  $N_H$   & $kT$ \\
& (10$^{22}\;$cm$^{-2}$) & (keV)\\
\hline
Best fit values  & 1.1 & 0.6 \\
Fixed temperature & 0.0--2.0 & 0.5 \\
Fixed absorption & 0.6 & 0.3--1.2 \\
\hline
\end{tabular}
\end{table}

\subsubsection{Comparison of the $Chandra$ observations\label{xExtent} \label{sect:pNumbers}}
A direct comparison of the photon numbers is not possible due to (a) different exposure times and (b) different detector responses. Nevertheless, we show in Fig.~\ref{fig:stripes} the mean photon number ignoring the differences in the sensitivity of the individual exposures. We note that all observations are compatible with this rough mean value using the error obtained by Gehrels weighting \citep{Gehrels_1986}.

For a more detailed comparison of the observations, a spectral model is constructed (not fitted) in each stripe. We experimented with models which use the overall best fit values and with models which reproduce the trend in the mean energies (see Fig.~\ref{fig:meanE}), they are listed in Tab.~\ref{tab:models}. As the predicted count numbers differ by less than one count, the statistical error overwhelms the error due to the unknown spectrum. 
Due to the different energy response of the 2001 ACIS-I and 2009 ACIS-S, the scaling factors relative to the 2005 observation are 0.83\dots0.87 and 0.91\dots1.20 depending on the assumed spectra (high photon energy and low photon energy, respectively).

The individual stripe models are normalized so that the total count number summed over the three ACIS observations in each stripe is conserved. Thus, the predicted total count number in a single stripe matches the observed value, which is statistically the best estimate for the model normalization for constant emission.
Figure~\ref{fig:stripes} exemplarily shows the result for the model with $kT=0.5\;$keV and variable absorption along the flow-axis (which is virtually indistinguishable from the model with constant absorption and variable temperature).

\subsubsection{Time variable emission?\label{sect:deviations}}
To test if the observations are statistically consistent with the hypothesis of constant emission, we perform Monte-Carlo simulations to estimate a confidence interval to accept or reject this hypothesis, i.e., to check if the observed photon distribution is an exceptional realization for a time constant emitting region. 
Since neither the time when new blobs appear nor their speed is predicted by current theories, the location of a new blob is not known a priori. Therefore, the statistical significance of any count number enhancement in a given region depends on the number of independent regions in which such an enhancement would be considered a knot. Essentially any stripe in Fig.~\ref{fig:stripes} can be regarded as a possible region for a knot, however, the result does not depend strongly on, e.g., the inclusion of stripes 1 or 7, although the general result does depend strongly on the set of stripes used. To ensure that our results are not biased by the particular selection of stripes, we repeated the simulation for different sets of stripes, which are two or three pixels wide and which have mutual offsets of one pixel.

One simulation involves a set of three new exposures simulated as new individual photon numbers for each stripe. These simulated photon numbers were based on the expected photon number in that stripe (sect.~\ref{sect:pNumbers}) and Poisson statistics. The individual likelihoods, i.e., the likelihoods to observe exactly the simulated photon number in a given stripe-observation combination, depend on the assumed expectation value for the photon number in this particular stripe. We derived this expectation value a posteriori from the simulated counts in a particular stripe using the relative efficiencies from sect.~\ref{sect:pNumbers}.
The total likelihood for each simulation was then calculated as the product of the individual likelihoods. Thus, the fraction of realizations with a total likelihood better than the observed one can be interpreted as the probability for time variability. This approach does not include background events (which is minor effect on the $\sim 1\%$ level) or other detector effects like alignment errors between the individual exposures.
We find probabilities between approximately 50\% and 96\% depending on the exact stripes used. As the spatial distribution of the luminosity along the jet is unknown, we cannot decide if the set of stripes indicating time constant emission or the set of stripes indicating variable emission, matches the real jet better. Thus, we conclude that time-variable \emph{and} time-constant emission are statistically acceptable.

A hypothetical moving, fading and X-ray emitting knot would result in a low photon number at the western end of the box during the 2001 observations since at that time the knot was located more closely to the driving source, a higher photon number during the 2005 observation, and virtually no photons by 2009 due to cooling and expansion.  The ``knot'' region (as indicated in Fig.~\ref{fig:overview}) shows exactly this count number pattern. Thus, such knots are also compatible with the observations.

In summary, the $Chandra$ X-ray observations can be regarded as statistically compatible with time constant X-ray emission. However, the significance is poorly constrained as it depends crucially on the stripes used.
The inability to find a clear sign for time dependence independent of the stripes used might be caused by the low number of counts since the 90~\% confidence level easily covers a range almost twice as large as the value itself in the outer regions of the outflow.  Therefore small scale time variability may be present, however, statistically time-dependent emission is not required.

\begin{figure}
  \centering
   \includegraphics[width=0.49\textwidth]{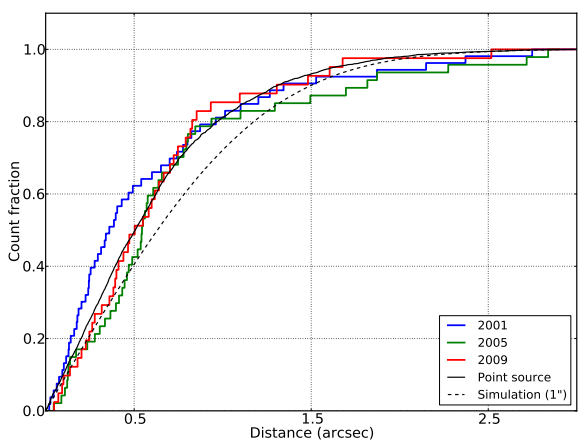}
   \caption{Encircled count fraction perpendicular to the jet axis using essentially the photons within stripes 2 and 3 of Fig.~\ref{fig:stripes}. The MARX simulation shows extended source for comparison.  \label{fig:yProjection} \label{fig:yProj}}
\end{figure}

\subsection{Extent perpendicular to the flow direction \label{yExtent}}
In order to check whether the $Chandra$ exposures show evidence for an extent of the emission perpendicular to the jet axis,  we have to take into account the exact position angle (PA) of the jet. Otherwise, the distribution of the photons around the jet axis would be artificially broader. We adopt a PA of 261$^\circ$ for the outflow \citep{Pyo_2009}.

As we know the absolute position of the jet axis only to about 0.5\arcsec, we estimate its position by the mean position of the photons perpendicular to the jet axis. The photon centroids along the N-S direction differ by about 0.14\arcsec{}, -0.18\arcsec{} and 0.04\arcsec{} from the mean position for the 2001, 2005 and 2009 observation, respectively, which is well within our estimated astrometric accuracy.

In order to check whether the X-ray emission region close to the driving sources is extended, we show in Figure~\ref{fig:yProj} the distribution of the photons perpendicular to the jet axis for the photons approximately in stripes 2 and 3 in Fig.~\ref{fig:stripes}. Due to the slight inclination of the jet axis with the x-axis, both regions do not overlap exactly. This figure also includes a  Marx\footnote{\texttt{http://space.mit.edu/cxc/marx/}} simulation of an extended (1\arcsec) source for comparison.
From this figure it is evident that the X-ray emitting region is smaller than 1\arcsec{}  and possibly smaller than 0.5\arcsec. We adopt a maximum extent of 0.5\arcsec{} for the lateral source extent but note that a source size smaller than this value is possible. We also note that the superposition of a number of smaller emission regions can mimic the observed photon distribution, thus making the physical extent of the emission region smaller than 0.5\arcsec.

The outer part of the emission region, approximately stripes 4 and 5, appear extended perpendicular to the jet axis. The lower number of counts prevents an estimate of the source size here but an extent on the 1\arcsec{} level is possible which would imply an opening angle of the X-ray emitting jet of about 7$^\circ$.

Figure~\ref{fig:yProj} shows slight deviations between the individual exposures, whether these are due statistical fluctuations or due to intrinsic changes in the emission region is hard to judge since variations are most evident on sub-pixel scales where, even with pixel-randomization turned off, the photon locations within the individual detector pixel become important. The standard deviation of the photon distances to the jet axis are 0.65, 0.69, 0.59 pixel (which is 0.492\arcsec) for the 2001, 2005 and 2009 observation, respectively (energy range 0.5-3.0 keV), i.e., rather similar values; the standard deviation for a point source is 0.57.
Therefore, we will concentrate in the following on changes of the plasma properties along the jet axis instead of changes perpendicular to it.

\subsection{Knot~D in X-rays? \label{sect:knotD}}
\begin{figure}
  \centering
   \fbox{\includegraphics[width=0.47\textwidth]{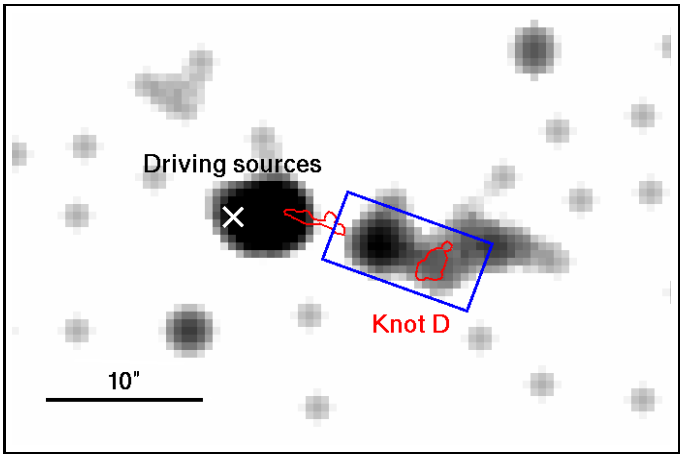}}
   \caption{Merged image of the region around HH~154 in the energy range 0.4\,-0.8\,keV. In order to show the diffuse emission, smoothed with a Gaussian kernel of 2.5\arcsec{}. The 2005 HST H$\alpha$ contours are shown. The box connects the inner H$\alpha$ emission with the one of knot~D and is of the same size as the box for the inner X-ray emission.\label{fig:knotD}}
\end{figure}

Between the outer part of the X-ray emission visible in Fig.~\ref{fig:stripes} and knot~D (or PHK~3), located at a distance of about 13\arcsec{} from the driving sources, excess X-ray emission is apparent (see Fig.~\ref{fig:knotD}). However, this excess is not clearly co-aligned with the expected postion of knot~D. The box shown in Fig.~\ref{fig:knotD}, which connects knot~D with the inner part of the H$\alpha$ emission, contains 11 photons in the energy range between 0.5\,keV and 3.0\,keV where only 4.5 background photons are expected. With a mean energy of 0.92\,keV, these photons are also softer than those closer to the driving sources.
Of these photons, two, three and six are from the 2001, 2005 and 2009 observation
as expected for a constant source due to the different efficiencies. There is not discernable spatial evolution. Assuming the absorption of knot~D also for the region around it ($A_V=2.5$ which implies $N_H=4.5\times10^{21}\,$cm$^{-2}$), the mean energy indicates a plasma temperature of about $3\times10^6$\,K. We caution that merging the observations covering about one decade can cancel out any physical structure and we regard the above as an indication that X-ray emission might be scattered around in this jet on a larger scale.

Whether the X-ray emission is indeed related to the jet, whether it persists over a decade, and what its true space velocity is cannot be answered from the data. However, we regard it unlikely that this X-ray emission represents knots originally heated much closer to the driving sources during the last decade as space velocities in excess of 1800\,km\,s$^{-1}$ are required to travel from the ``knot'' region (cf.,~Fig.\ref{fig:overview}) to the position of knot~D within four years. The apparently knot like structure in Fig.~\ref{fig:knotD} contains at most two photons from the 2009 observation while three are from the previous observations. In any case, the center of this structure is located about 6\arcsec{} from the ``knot'' region so that space velocities in excess of 1000\,km\,s$^{-1}$ are still required to travel from the ``knot'' region to this location. Therefore, this X-ray emitting plasms might be the remnant of jet material emitted more tha a decade ago.

\section{Discussion\label{sect:disc}}
The interpretation of the X-ray findings depends crucially on the required mass loss rate, shock velocity, plasma density and the cooling time of the X-ray plasma within the flow. Therefore, we start our discussion by deriving estimates for their respective values.

\subsection{Mass loss rate and shock velocities}
Using the formula given in \citet[][eq.~2]{Schneider_2008} for the massloss rate required to explain the observed X-ray emission we find $\dot{M}_{X-ray} \approx 1.5\times10^{-12}\,M_\odot\,\mbox{yr}^{-1}$ or approximately a factor of 10 lower than the inner X-ray emission component of DG~Tau. This formula assumes that the material observed with $Chandra$ is shocked only once, which should be a relatively good approximation since high shock velocities are required to heat the material above $T\gtrsim10^6\,$K. According to \citet{Raga_2002}, shock velocities of approximately 700~km\,s$^{-1}$ are needed to heat the material to the observed 0.6~keV close to IRS~5, while only $v_{shock}\approx$550~km\,s$^{-1}$ is needed for the lower temperature at larger distances in case this outer X-ray emitting plasma is heated in situ and not the cooled down remnant of the inner X-ray component.

\subsection{Densities\label{sect:dens}}
The low densities of the HH flows cannot be directly measured with X-rays, however, we can derive a lower limit on the density assuming a certain emission volume. From our analysis of the source extent in sect.~\ref{xExtent} and sect.~\ref{yExtent}. we assume that the emitting volume is extended perpendicularly  to the jet axis by 0.5\arcsec$\hat{=}$70~AU and along the jet axis by 2\arcsec{} and estimate that at least 80~\% of the photons originate in this region. These estimate translates into a maximum volume of $5\times10^{45}\;$cm$^{-3}$ for an inclination angle of 45$^\circ$. For a filling factor of unity, the electron density of X-ray emitting plasma is then
\begin{equation}
n_X=\sqrt{\frac{EM}{0.85\cdot V}}\approx 1.2\times10^3\;\mbox{cm}^{-3}.
\end{equation}
This value is a lower limit on the density since the plasma might be concentrated in individual denser clumps, i.e., the volume filling factor could be less than unity. Very close or co-spatial to the X-ray emission material of lower temperatures ($\sim10^4\;$K) has been observed (e.g. [Fe II]), so we consider a filling factor of unity rather unlikely since some intermediate temperature material will connect both temperature components. The lower limit of the electron density in the outer part of the X-ray emitting jet is lower by a factor of about four due to the decreasing surface brightness assuming an opening angle of 0$^\circ$ and no change in the plasma properties\footnote{In sect.~\ref{sect:eTrend} we argue that the absorbing column density and the temperature decrease along the outflow. This changes the lower limit of the density in the outer parts of the X-ray emitting jet by a factor of 1.4.}.

In the optical and near-IR direct density values from line ratios have been derived mainly for the outer part of the jet where densities of a few $10^3\,$cm$^{-3}$ are found, e.g., the [Fe~{\sc ii}] lines imply $n_e=7.6\times10^4\;$cm$^{-3}$ \citep{Liseau_2005}. Densities up to $10^6\,$cm$^{-3}$ have been derived for the inner 2\arcsec~ by \citet{Itoh_2000} from an analysis of [Fe {\sc ii}] near-infrared lines. Note that the [Fe {\sc ii}] lines have a higher critical density than the [S~{\sc ii}] lines usually used for the density measurement. High hydrogen densities have also been found close to the driving sources for a few other sources \citep[e.g.][]{Melnikov_2009, Bacciotti_2000}.

Considering the thermal pressures of these two temperature components, we find that the lower limit on the thermal pressure of the X-ray emitting plasma ($T_X \approx 7\times10^6\,$K) is
\begin{equation}
P_X=2\,n_{X}k_BT_{X}\approx2\times10^{-6}\;\mbox{dyn}/\mbox{cm}^2\,.
\end{equation}
The high densities of the material observed in [Fe~{\sc ii}] results in an approximate thermodynamic pressure equilibrium of both components, i.e., densities of a few times $10^6\,$cm$^{-3}$ suffice to provide the required pressure at a temperature of $T\approx10^4\,$K.
This high electron density close to the driving sources supports the idea that the density decreases by approximately two orders of magnitude within the innermost 5\arcsec$\hat{=}$1000~AU (deprojected) as estimated from the [Fe~{\sc ii}] and [S~{\sc ii}] lines.

A conical outflow decreases its density by exactly two orders of magnitude from 0.5\arcsec{} to 5\arcsec{} for a constant outflow velocity. However, HH~154 is likely not strictly conical since \citet{Fridlund_2005} noted that the opening angle close to the driving source might be as large as 90$^\circ$, which is consistent with the estimated lateral jet size of 0.5\arcsec{} by \citet{Pyo_2002} at this distance. Therefore, the density decrease beyond 0.5\arcsec{} is probably less than for a conical outflow as an opening angle of only about 3$^\circ$ (see footnote \footnote{The size of the Mach-disk of knot~D located at a distance of $\sim$10\arcsec{} \citep[0.6\arcsec,][]{Fridlund_2005} indicates the local size of the jet.}) does not suffice to decrease the density sufficiently for large inital opening angles.
Thus, it is not clear if or where the ``cold'' jet component is in pressure equilibrium with the X-ray emitting plasma.

Another possibility is that the magnetic pressure supports the X-ray emitting volume against expansion. We estimate its strength by assuming a plasma-$\beta$ of unity ($\displaystyle B^2= 8\pi P_{gas}$) and find  $B\approx 6\, \mbox{mG}\,$.
Such a value is reasonably close to the driving source \citep[see Tab.~1 in ][]{Hartigan_2007} and requires lower densities than pure pressure support. The lower limit would still be $n_e\gtrsim4\times10^4\,$cm$^{-3}$ if the magnetic field scales with the density as expected ($B\sim n^p$ with $p=0.5\dots1$) since measured magnetic fields in HH objects indicate 15\,$\mu$G for $n=100$\,cm$^{-3}$ \citep{Hartigan_2007}.  The interstellar magnetic field seems too weak to collimate the jet \citep{Cabrit_2007}, however, MHD self-collimation is a likely scenario. Therefore, the same wound-up (helical) magnetic field which collimates the outflow can lend support for the X-ray emitting plasma, largely inhibiting lateral expansion of the hot X-ray emitting material.

\subsection{Plasma cooling \label{sect:cooling}}
Three processes contribute to the cooling of a plasma: Radiative cooling, cooling by expansion and thermal conduction.
The pressure work done by the plasma is $\delta W = p dV$ and the radiative losses are
\begin{equation}
\delta Q_{rad} = -n_e^2 V(t) \Lambda(T) dt\,,
\end{equation}
where $n_e$ is the electron density, $V$ the volume and $\Lambda(T)$ is the cooling function.
The conductive heat flux is given by
\begin{equation}
q_{cond} = -\kappa(T)\bigtriangledown T\,, \label{eq:cond}
\end{equation}
where the thermal conductivity according to Spitzer is
\begin{equation}
\kappa(T) = \kappa_0 \frac{T^{5/2}}{\ln \Lambda}\;\mbox{erg}\,\mbox{s}^{-1}\,\mbox{K}^{-1}\,\mbox{cm}
\end{equation}
with $\kappa_0=1.8\times10^{-5}$ and the Coulomb logarithm $\ln \Lambda$, which describes the collision properties of the plasma and is of order 10.
When the mean free path length for energy exchange is of the same order as the thermal scale height, the conduction should be approximated by the saturated flux
\begin{equation}
q_{sat} = 5 \phi \rho c_s^3\,,
\end{equation}
with $\phi\approx0.3$ \citep[e.g.][$\rho$ is the mass density and $c_s$ is the local sound speed]{Borkowski_1989}. For an estimate of the importance of the saturated flux, we assumed a linear temperature decrease. Under these circumstances the Spitzer value  exceeds the saturated flux on spatial scales of about 10\,AU for the cooling from $T_1=10^6$~K to $T_0=10^4$~K, i.e., the saturated flux should be used for these steep gradients ($n\approx10^3$).

Thermodynamics states that the energy change of a plasma cell is described by
\begin{equation}
dU + \delta W = \delta Q = \delta Q_{rad} + q_{cond}\cdot A\;dt\; \label{eq:cooling1}
\end{equation}
with the internal energy $U=\alpha N kT$ ($\alpha=3/2$ for a fully ionized plasma), the particle number $N$, the Boltzmann constant $k$, the temperature $T$  and $A$ is the surface area through which heat conduction proceeds.
We use $p=2n_ekT$ in the expression for the pressure work and follow \citet{Guedel_2008} by writing  eq.~\ref{eq:cooling1} as
\begin{equation}
\alpha \frac{dT}{T(t)} + \frac{dV}{V} = - \left( \frac{n_e \Lambda(T)}{2 k T(t)}+ \frac{\kappa_0}{2n_eVkT}\cdot A\frac{T^{5/2}}{\ln \Lambda} \nabla T \right) dt \,,\label{eq:cool}
\end{equation}
where we used $N_e=n_e\cdot V$ and note that this expression holds only in the presence of sufficiently small temperature gradients.

In order to estimate the relative importance of the three cooling terms, additional information is needed, in particular, the opening angle of the X-ray emitting jet, its density structure, the temperature gradient, the surface for the heat conduction, which would include the magnetic topology and the properties of the environment, e.g., its ionization. These quantities are not available for the X-ray emitting part of the jet. We therefore decided to give some order of magnitude estimates for the cooling times of the individual processes ignoring contributions of the other ones. As we will see, there are distinct regions in the parameter space where each process seems to dominate, so we regard this approach reasonable.

\begin{figure}
  \centering
   \includegraphics[width=0.49\textwidth]{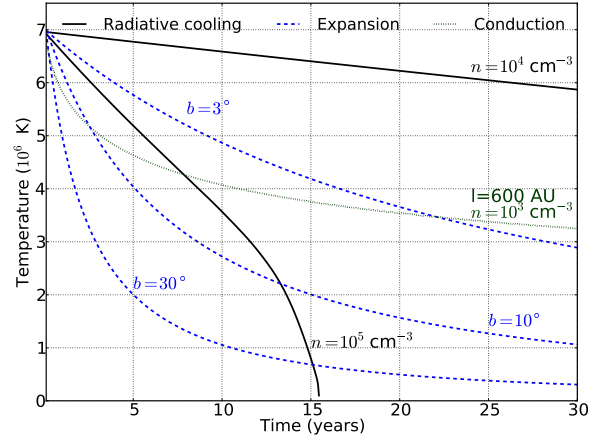}
   \caption{Cooling curves for the different cooling processes. Parameters of the models are labeled (density: $n$, opening angle: $b$, cooling length: $l$). \label{fig:cooling}}
\end{figure}

Figure~\ref{fig:cooling} shows the cooling curves for the different processes assuming different parameters for the jet. For radiative and conductive cooling, the mapping of time to distance in this figure depends on the actual, deprojected space velocity of the plasma. 
A rough estimate is 0.3\arcsec{}~yr$^{-1}$, which implies that today's inner X-ray emission will reach the position of today's outer emission in 15 years.
Adiabatic cooling, on the other hand, does not depend of the outflow velocity but only on the initial cross-section of the plasma and on the opening angle.

\subsubsection{Adiabatic cooling\label{sect:aCooling}}
Protostellar jets usually show an approximately conical structure at some distance from the driving sources so that the flow expands mainly perpendicular to the jet axis.
In the limit of adiabatic cooling
\begin{equation}
T\,V^{\gamma-1} = \mbox{const}
\end{equation}
holds. Since we do not observe local temperatures, we have to average the temperature, weighted by density\footnote{Note that $EM=n^2\,V=n\,N$ with a constant number of particles $N$ in each cell.}, over the volume used to measure the temperature. We use the following approximation for the volume of the plasma cell
\begin{equation}
V=\pi l\left(r_0+r\tan b \right)^2\,,
\end{equation}
where $r(t)=v\cdot t$ is the position along the jet axis measured from the initial distance, while $r_0=0.25\arcsec$ is the initial jet radius at this position and $2\cdot b$ the opening angle ($l$ is the length of the cell along the jet axis).
The initial cross-section is fixed and the temperature decrease depends only on the position along the outflow. 
From the size of the Mach disk about 10\arcsec~ from the driving sources, we estimate an opening angle of $3$\,--\,$10^\circ$ for the flow, where the separation of the working surface and the Mach disk argues for values closer to $3^\circ$.
Different outflow velocities would change the curve for the expansion cooling in Fig.~\ref{fig:cooling} but would not lead to another spatial temperature structure, because the dependence on $v$ cancels out in the equations.

As described by \citet{Guedel_2008}, the expansion additionally reduces the density of the emitting plasma and thereby lowers the number of emitted photons more strongly than expected on the basis of the temperature decrease alone. For a consistency check, we calculated the expected number of photons at 3.5\arcsec~ from the driving sources from the ratio of the emission measures at 0.5\arcsec~ and 3.5\arcsec~ and the drop in temperature. Assuming constant absorption, we expect a drop in photon number by approximately a factor of about 6 from 0.5\arcsec~ to 3.5\arcsec~ for an opening angle of 3$^\circ$, which is approximately compatible with the observed value. The larger opening angle of 10$^\circ$ would reduce the photon number more strongly, i.e., the combination of the temperature and  density decrease reduces the expected photon number by about 200 for the same distance.

\subsubsection{Radiation cooling}
We solved eq.~\ref{eq:cool} using the cooling function of Chianti version~6.0 \citep{Dere_1997,Dere_2009} assuming half solar metallicity.
Figure~\ref{fig:cooling} shows two cooling curves for radiative cooling. According to eq.~\ref{eq:cool}, the cooling time depends linearly on the density. It is clear that radiative cooling does not contribute significantly to the cooling as long as the density does not exceed $n\approx10^4\,$cm$^{-3}$.

\subsubsection{Conductive cooling}
Magnetic fields are essential for the launching of jets, but even at greater distances, small magnetic fields ($\sim 100\,\mu$G) influence the jet dynamics \citep[][]{Hartigan_2007}. They can also strongly suppress heat conduction perpendicular to the field lines even for weak fields \citep[$\sim 1\mu\;$G, see eq. 5-53 in ][]{Spitzer_1962}. In the presence of turbulent magnetic fields, heat conduction might be suppressed by about two orders of magnitude or even enhanced relative to the Spitzer value \citep[e.g.][]{Narayan_2001,Cho_2003,Lazarian_2006} depending on the scale of the turbulence. We regard it as plausible that heat conduction works most efficiently along the jet axis while it is suppressed by some kind of magnetic field 
perpendicular to the jet axis. The Spitzer value for the heat conduction assumes an ionized plasma, which might not be entirely true throughout the jet, however, a considerable amount of ionized material should be present close to the X-ray emitting plasma.
Given these uncertainties, we estimate conductive cooling by 
\begin{eqnarray}
\tau&=&2.6\times10^{-9}\frac{n l^2}{T^{5/2}} \,\mbox{s}\\
    &\approx& 52 \left(\frac{n}{1000\,\mbox{cm}^{-3}}\right) \left(\frac{l}{210\,\mbox{AU}}\right)^2 \left( \frac{T}{3\times10^6\,\mbox{K}}\right)^{-5/2}\,\mbox{years} 
\end{eqnarray}
given in \citet{Orlando_2005}. We show in Fig.~\ref{fig:cooling} a cooling curve by numerically integrating the conductive cooling for a fixed density $n$ and for
cylindrical geometry ($V\approx A\cdot l$).
The effect of the conductive cooling depends on the density of the plasma and on the temperature gradient, i.e., on the cooling length (we assumed 600\,AU for a temperature decrease from 0.6\,keV to 0.1\,keV). The curve shown in Fig.~\ref{fig:cooling} is intended to give a rough impression of this effect and we caution that the provided estimate for the conductive cooling might be off by orders of magnitude in some scenarios, e.g., for turbulent magnetic fields.

\subsubsection{Cool conclusions}
From Fig.~\ref{fig:cooling} it is clear that cooling by expansion dominates over radiation. Whether conduction is important depends on the density, the temperature gradient and the magnetic field configuration. When no heat is transferred perpendicular to the jet axis, we expect adiabatic cooling to dominate. We will therefore focus on that cooling process in the following.

\subsection{Trend in mean energy \label{sect:eTrend}}
Judging from the absorption value ($N_H\approx1.4\times10^{22}\;$cm$^{-2}$) close to the driving sources, the visual absorbing magnitude is approximately $A_V\approx8\;$mag \citep{Vuong_2003}. \citet{Itoh_2000} analyzed the [Fe~{\sc ii}]~1.644$\;\mu$m/1.257$\;\mu$m line-ratio as an estimate for the evolution of the visual extinction along the jet axis and found a value of $A_V\approx7$ for distances greater than 1\arcsec{} from IRS~5. This is compatible with our estimate from the X-ray spectrum. Therefore, we expect that the correlation of $A_V$ and $N_H$ holds in the jet region. Closer to the driving sources the extinction increases up to $A_V\approx21$. \citet{Fridlund_2005} also derived $A_V$-values, which are slightly lower than the values of \citet{Itoh_2000} for the optical knots located further downstream. They find that the extinction decreases slowly towards the outer knots where $A_V\approx2-3$ is found. At the position of knot~F which is closest to the driving sources, they estimate an absorption of $A_V>4\,$.

Assuming that the absorbing column density decreases from $N_H=1.4\times10^{22}\,\mbox{cm}^{-2}$ at 1\arcsec{} from the driving sources to $N_H=8\times10^{21}\,\mbox{cm}^{-2}$ at 5\arcsec{}, the plasma temperature still has to decrease from 0.7~keV to 0.4~keV in order to explain the decrease in the mean energy. We therefore conclude that the temperature decrease of the X-ray plasma significantly contributes to the  softening of the photons beyond $\approx1$\arcsec. 
It is possible that some emission is coming from the innermost 1\arcsec{} where it is more strongly absorbed. 
For an upper limit on the temperature change along the flow axis, we fix the absorption to $N_H=8\times10^{21}\;$cm$^{-2}$, which requires, according to Fig.~\ref{fig:meanE}, that the plasma temperature decreases from approximately 1.7~keV to below 0.2~keV within~3\arcsec{}. 

An opening angle of a few degrees reduces the plasma temperature along the outflow as required by the above estimates, consistent with the assumption that adiabatic cooling dominates the plasma cooling. As discussed in sect.~\ref{sect:aCooling} the reduction of photons along the jet axis is compatible with a shallow opening angle.

As we have no good estimate for the cooling time of the plasma, it is possible that cooling is very efficient and re-heating of the X-ray emitting plasma along the inner few arcsec is required in order to produce observed X-ray emission at larger distances from the driving sources. However, the temperature decrease along the jet axis remains virtually constant over 10 years of observation so that also the hypothetical re-heating must be relatively constant over this period of time. Large individual blobs with largely varying velocities, ejected every few years, would probably produce a more variable temperature structure. Smaller, unresolved internal shocks could, on the other hand, be present so that the cooling time of the plasma can be shorter than derived from the decrease in the mean energy.

\subsection{The inner emission component}

The most striking feature observed in all observations is the existence of a luminous X-ray emission region close to the
driving source(s). The peak of this feature is approximately 0.5-1.0\arcsec{} offset from L~1551~IRS~5 (R.A.(2000)=$04^\text{h}31^\text{m}34$\fs 15, Decl.=18\fdg08'05\farcs04) towards the south-west.
Its luminosity and temperature remains virtually constant within a timespan of about nine years. 

Therefore, it seems necessary to review the arguments which prohibit an association of this inner component with one or both of the central driving sources themselves. Essentially, these are (a) the astrometry and (b) the absorption. Concerning (a), the centroid of the inner component is placed at least 0.5\arcsec{} from the location of radio sources in every of the three available $Chandra$ observations. Although this is marginally compatible with our estimated astrometric accuracy, this is unlikely to be caused by a repeated incorrect pointing of the satellite since the centroids (using the photons in the inner $r=1$\arcsec{} circle) of the inner component match to within 0.3\arcsec{} for all observations.
As to (b), the interpretation of the scattered light and the non-detection of direct emission places a firm lower limit of \mbox{$A_V \gg 20$} on the absorption towards IRS~5 \citep[][]{Stocke_1988}. This translates into an absorbing column density of $N_H \gg 4\times10^{22}\;$cm$^{-2}$, at least three times higher than measured. The measured value, on the other hand, corresponds well to the one estimated for the inner jet from the near-infrared line-ratios and suggests that X-ray and near-infrared emission spatially coincide.

\citet{Bally_2003} sketched a possible scenario in which scattered X-rays are responsible for the observed X-ray emission. However, these authors concluded that this option is less likely and we agree with their evaluation. The parameters required for this scattering scenario, e.g., densities of $n\sim10^9\,$cm$^{-3}$, are not strictly ruled out, but would be exceptional for protostellar jets. Furthermore, the trend in the mean energy also argues against the scattering scenario, because Thomson scattering is not very sensitive to the scattering angle and independent of the wavelength. Dust scattering, on the other hand, is by far not sufficient to explain the observations using the usual conversion factors.

We therefore associate this feature with the apparently stationary [Fe~{\sc ii}] emission complex observed by \citet{Pyo_2009, Pyo_2005, Pyo_2002}. Their [Fe~{\sc ii}]~$\lambda1.644\;\mu$m data, obtained over a timespan of four years, show an apparently stationary component close to the driving sources. \citet{Pyo_2009} already proposed that the inner X-ray emission component is associated with the innermost [Fe {\sc ii}] emission peak, called PHK1 (distance to IRS~5: 1.1\arcsec{}).
The total flux in [Fe~{\sc ii}] ($\sim5\times10^{-15}\;$erg/s/cm$^2$) is within an order of magnitude comparable with the X-ray flux (0.5-10.0\,keV: $4\times10^{-15}\;$erg/s/cm$^2$ or unabsorbed $3\times10^{-14}\;$erg/s/cm$^2$).

The [Fe~{\sc ii}] emission at PHK1 is dominated by low velocity material with $v\sim-60\dots-150\;$km/s (or deprojected $v=-85\dots212\,$km/s, for an inclination of $i\approx45^\circ$). Interestingly, the post-shock velocity of a shock with an initial velocity of v=700~km/s is 175~km/s, i.e., within the range of the low velocity component.
\citet{Pyo_2002,Pyo_2009} noted that the velocity dispersion of the low velocity [Fe~{\sc ii}] emission decreases with increasing distance to the driving source, which they interpret as a collimation of the outflow. The large opening angle of the flow close to the driving sources and the shallow opening angle further downstream support this interpretation. Collimation might also be responsible for the X-ray production and would naturally explain their stationary appearance. Since the high-velocity [Fe~{\sc ii}] material appears approximately where the X-ray emission disappears, it is tempting to associate this material with outflowing plasma not as strongly shocked as the X-ray emitting material. However, the total mass-loss derived for the X-ray component is lower than for the optical part of the jet so that it remains unclear whether the ``absence'' of highly blueshifted emission close to the driving sources is somehow connected to the existence of X-ray material, i.e., if a large fraction of shocked high-velocity material reaches X-ray emitting temperatures.

\subsection{The extended or outer emission component \label{sect:oDisc}}
During the 2005 ACIS-I observation, an enhancement of photons two arcseconds downstream from the bulk of the X-ray emission is evident \citep{Favata_2006}. We estimated in sect.~\ref{sect:deviations} that the low number of counts in the corresponding region might be a statistical fluctuation. Nevertheless, it is still possible, if not even physically plausible, that the elongation differences are caused by a transient X-ray emitting knot, possibly comparable with other X-ray emitting knots within HH objects.

Concerning the position of this blob, its 2005 position coincides with one or all emission peaks in the ``F''-complex \citep{Bonito_2008}. The space velocities measured in this region range from $\sim100\;$km$\,$s$^{-1}$ to 500$\;$km$\,$s$^{-1}$ in optical forbidden emission lines \citep[][]{Fridlund_2005}. At this distance from the driving source the high-velocity component in [Fe~{\sc ii}] becomes dominant over the low-velocity component, and velocities up to 500~km$\,$s$^{-1}$ have been measured \citep{Pyo_2005}. Furthermore, \citet{Pyo_2005} noted that the outer high velocity [Fe~{\sc ii}] component might exhibit time-variability at approximately the same time of the appearance of the X-ray knot. However, the F-complex did not change much during this time in its optical appearence \citep{Bonito_2008}.

The distance traveled by this hypothetical knot between 2005 and 2009 would be approximately 0.6-3.0\arcsec{} (100~km$\,$s$^{-1}$ \dots 500~km$\,$s$^{-1}$). In the corresponding regions zero or one photon are recorded during the 2009 ACIS-S exposure, which is more sensitive than the previous ACIS-I observation at low energies.
A maximum of seven photons can be attributed to the 2005 ``knot'' (cf. Fig.~\ref{fig:overview} where the most favorable geometry is sketched), so that the luminosity of this knot must have decreased during the last four years if it indeed existed.
For the interpretation of these phenomena, the cooling time of the X-ray emitting plasma is crucial but unfortunately not known with the required precision.

When the cooling time of the hot plasma is short compared to the travel time to the outer locations of X-ray emission, it is impossible that the material is only heated  close to the driving sources and then just cools while it is flowing outwards. Internal shocks are a natural explanation for the re-heating, but the observations require a nearly constant decrease in shock velocity with increasing distance to the driving sources (see the trend in the mean energy, sect.~\ref{sect:eTrend}).

If, on the other hand, the plasma is not significantly re-heated while flowing outwards (no internal shocks), either a variable mass outflow or a statistical fluctuation are responsible for the apparent knot. In any case, a rather constant temperature close to the driving sources is required, which translates to a constant shock velocity for shock heating.
In both cases, a shallow opening angle is mandatory and magnetic fields probably suppress heat conduction efficiently.
The decrease in plasma temperature along the flow reflects the cooling time of the plasma and explains, why no emission is detected at larger distances from the driving sources.

\subsection{Comparison with DG Tau and other jet X-ray sources}
The so-called TAX sources \citep{Gudel_2007}, which show spectra composed of two emission components with vastly different absorbing column densities, e.g. from an embedded star and a less embedded jet, show a striking similarity to HH~154. We associate the beehive proplyd and similar COUP sources with this group \citep{Kastner_2005}.

The comparison of HH~154 with the X-ray emitting jet of DG~Tau \citep{Guedel_2005, Gudel_2007, Guedel_2008, Schneider_2008, Guenther_2009, Guedel_2011} is particularly interesting as multi-epoch high spatial resolution X-ray observations are also available. 
The important similarity between both jets is that the majority of the X-ray emitting plasma is located close to the driving source(s) in all observations. We will therefore focus our comparison on this innermost part of the outflow, the resolved outer part of DG~Tau's jet is probably more related to the outer part of HH~154 (knot~D, see sect.~\ref{sect:knotD}). The inner part of the outflow also appears qualitatively similar in the [Fe II] position-velocity maps \citep[e.g.][]{Pyo_2003,Pyo_2009}. Both outflows exhibit a low velocity component close to the star, while the high velocity component is located further downstream. Other jet-driving sources like HL Tauri and RW Aurigae show a slightly different pattern in [Fe II] \citep{Pyo_2006} with the low velocity component at a larger distance from the stellar position.

The cooling times of the X-ray emitting plasma, however, appear to differ between the two jets. While we estimated in sect.~\ref{sect:eTrend} that the temperature gradient along the outflow indicates a cooling distance of the X-ray emitting plasma in the range of a few 100\,AU., the essentially unresolved inner component of DG~Tau \citep{Schneider_2008, Guedel_2011} implies a much shorter cooling time for DG~Tau. This might be related to the higher minimum electron density of the X-ray emitting plasma of this component in DG~Tau ($n_e>10^6\,$cm$^{-3}$) compared to HH~154 ($n_e>1.2\times10^3\,$cm$^{-3}$). While the X-ray emitting plasma of HH~154 can be in pressure equilibrium with the cooler jet material, this seems much less likely for the X-ray emitting jet of DG~Tau. 
The shorter distance of the inner X-ray jet component of DG~Tau to its driving source compared to HH~154 might be the cause for the higher density of DG~Tau's jet X-ray emission. Assuming that the heating happens more closely to the driving source for DG~Tau's jet, a large opening angle of the outflow can explain the essentially point-like appearence of the X-ray emission in DG~Tau as cooling by expansion would reduce the plasma temperature sufficiently fast. The variable X-ray luminosity without measurable differences in the position of the inner X-ray jet component \citep{Guedel_2011} also argues for a short cooling distance of this X-ray component as otherwise at least shifts in the mean position of the X-ray emission should be measured assuming the proper motion of the optically observed knots ($\sim0.3\arcsec$\,yr$^{-1}$). For HH~154 a shallower opening results in a larger extension of the X-ray emission. One can speculate that the X-ray emitting plasma contributes to the expansion of the jet close to the driving source in DG~Tau while that is not as clear for HH~154. 

An obvious difference between the X-ray jets of DG~Tau and HH~154 is the different plasma temperature close to the driving source. However, the soft component of the beehive proplyd has approximately the same plasma temperature as HH~154 \citep{Kastner_2005}. For DG~Tau, the inner X-ray component is only marginally hotter than the outer, resolved one at a distance of 5\arcsec{} from DG~Tau, while for HH~154 the outer parts of the jet (e.g. knot~D or the outer stripes in Fig.~\ref{fig:stripes}) seem to be cooler by a factor of two compared to the emission close to the driving source. Since adiabatic cooling decreases the temperature strongly along the flow, it seems unlikely that the same plasma found close to the driving source is also responsible for the outer X-ray component of the DG~Tau's jet or knot~D in HH~154, and additional internal shocks cause the high temperatures further outwards \citep[e.g. as in the models of][]{Bonito_2010a}. 

\citet{Stelzer_2009} recently detected the appearance of an X-ray knot after an FU~Ori like outburst of Z~CMa, i.e., X-ray emission located about 2000~AU from the driving source. This emission is much farther out than in HH~154, which can either indicate that the lifetime of such knots might be relatively long or that strong shocks also occur further outwards in the flow.
As in Z~CMa the X-ray emission in other HH objects is located further out along the jet. If the correlation of the X-ray emission region with one of the working surfaces in these outflows is an observational bias, such that X-rays between the knots are less likely to be recognized as associated with the HH object or if it is an intrinsic feature of the X-ray production mechanism is currently not clear. 

In summary, two properties make X-rays from the protostellar jets identifiable, (a) a TAX-like spectrum and (b) X-ray emission displaced with respect to the driving source. Strictly speaking, HH~154 belongs to the TAX class of objects and the strong absorption of the driving sources makes the detection of extended emission possible. We speculate that the X-ray morphology of HH~154 also applies to the other TAX sources where the angular resolution is insufficient to resolve all details, e.g. due ``contamination'' by stellar emission.
As the outflow rate required to produce the observed X-rays in these objects is lower than estimated from the optical emission by a few orders of magnitude, it seems likely that such a high velocity component has escaped detection in other wavelength regimes, but is still powerful enough to lead to the observed X-ray fluxes.

\section{Model implications\label{sect:models}}
After the discovery of X-ray emission offset from the driving sources (IRS~5) by \citet{Favata_2002} and \citet{Bally_2003}, a variety of models have been proposed in order to explain this phenomenon. In particular \citet{Bally_2003} described an ensemble of possible explanations covering a broad range of possibilities including, e.g., X-rays from the driving source(s) reflected into the line of sight from the outflow cavity. \citet{Bonito_2010b} performed detailed magneto-hydrodynamical simulations of a jet with a variable outflow velocity focusing on  high velocity shocks within the outflow and the associated X-ray emission. These authors discuss four different scenarios for the evolution of the X-ray emission of HH~154, one of which considers a stationary source. 

We concentrate on models related to the apparently stationary X-ray emission complex since the majority of the X-ray emission is related to this complex. The scenarios related  to the interpretation of the outer or extended X-ray component were discussed  in sect.~\ref{sect:oDisc} and depend crucially on the unknown cooling time of the X-ray emitting plasma.

\subsection{A jet with random ejection velocity/Internal shocks}
This is the model discussed by \citet{Bonito_2010b,Bonito_2010a}. Their simulations can produce an emission complex close to the driving source, when a recently ejected faster blob overtakes a more slowly moving blob (cf. their Fig.~2). In the absence of strong cooling, the proper-motion of such a knot would be detectable with the available high resolution X-ray observations. 
Concerning the inner, apparently stationary source, the models of \citet{Bonito_2010a} predict that the most probable position of a shock is close to the driving sources.

The virtually constant X-ray luminosity and the relation to the constant [Fe II] emission argues against a strong variation of the shock velocity or location. One solution for these discrepancies is a relatively regularly modulated jet so that a constant luminosity might be mimicked by the superposition of a roughly constant number of smaller shocks formed close to the driving source. The trend in the mean energy would then reflect the cooling of these smaller individual blobs, while they travel along the outflow \citep[e.g., see also the sub-radial blobs modeled by ][]{Yirak_2009}.
Another possibility is that variations in the shock properties are hidden by the low photon numbers and the inner X-ray emission is caused by larger knots shocked close to the driving source. The constant appearance would then not reflect a constant outflow but would rather be a chance coincidence.

In this model the absence of X-ray emission farther downstream would be explained either by a lower density or a low temperature of the plasma inhibiting its detection. The opening angle of the optical jet is roughly consistent with this picture. It requires, however, that strong shocks at larger distances from the driving source are  less probable than close to the driving sources, which is true for the models of \citet{Bonito_2010b}. In case of a short cooling time the decrease of the plasma temperature reflects a decreasing shock velocity with increasing distance to the driving sources, which would provide another explanation for the non-detection of X-ray emission farther downstream. In any case, the observations clearly show that the heating to X-ray temperatures is a function of the position along the flow.

\subsection{Base/collimation shock}
Guided by the first $Chandra$ observation, \citet{Bally_2003} proposed that some kind of stationary base-shock can explain the observed X-rays, either independently for each driving source or at the envelope of both sources. 
In these scenarios,  the magnetic fields can collimate the outflow and can also support the jet against the thermodynamic pressure of the hot X-ray emitting plasma.

This scenario requires lower velocities than the internal shock model, but still higher than detected in available spectra. The ``deflection'' angle might be relatively large ($\sim 45^\circ$) as the opening angle close to the driving sources might also be large so that flow velocities of $10^3\;$km\,s$^{-1}$ suffice for the X-ray production.
Also, the concentration of the X-rays within a rather small volume close to the driving sources and the virtually constant X-ray luminosity are a natural consequence of this scenario. The base-shock scenario does not inhibit jet mass flux variations and is consistent with the observations as long as the amplitude of these variations is small enough. For sufficient clumpiness and blob ejection cadence, the base-shock model and the internal shock model become indistinguishable and share the possibility for small amplitude time variability.

Concerning the location of the base-shock, it seems pleasing to attribute the brightest X-ray spot to the location of the base-shock consistent with the [Fe~{\sc ii}] observations of \citet{Pyo_2009}. However, the increasing absorbing column, which seems to cause the hardening of the photons in the innermost part of the flow, can absorb the soft X-rays closer to the driving sources. Thus, it is possible that the true location of the shock region is hidden behind a larger absorbing column and located closer to the driving source.

The trend in the mean energy is also a natural outcome of this scenario, when the plasma is heated to X-ray emitting temperatures close to the driving sources and cools while flowing outwards. Adiabatic cooling, providing an upper limit on the cooling time, is approximately consistent with the observed trend in the mean energy.

\subsection{Precessing jet}
Jet precession seems to be required for some of the observed jets \citep[e.g.~HH~34][]{Masciadri_2002}. The precession times are usually rather long ($\gtrsim 10^3\,$yrs), therefore, the change in outflow direction for HH~154 would be small between the 2001 and the 2009 $Chandra$ observations. Still, some kind of a drilling effect might be present. A constant flow hitting different parts of the envelope would lead to a constant appearance of the inner emission component. As the opening angle ($\sim90^\circ$) close to the source is probably large compared to the expected precession angle, we regard it as less likely that the direction change of the outflow is responsible for the X-ray emission.

\subsection{Stellar wind}
The solar wind has roughly the temperature and velocity observed for HH~154's X-ray emission. We can imagine that during the early stellar evolution the outflow rates of stellar winds are much higher than for the present-day Sun and that the same process leading to the collimation of the slower outflow components collimates the stellar wind. A stellar wind might be important for the angular momentum problem. but cannot be responsible for outflow rates above $10^{-9}\;M_\odot/$year due to the resulting excessive X-ray emission \citep{Decampli_1981,Matt_2008}. However, the outflow-rate required for the observed X-ray emission is orders of magnitudes lower and \citet{GdC_2007} found evidence for a stellar driven wind for RY~Tau. These authors suggest that the superposition of many individual  small-scale outflows from the stellar surface leads to observed morphology of the FUV~lines. Therefore, a stellar wind, while not responsible for the main outflow, might provide the required high temperature plasma close to the driving source. This stellar wind would not require high outflow velocities for the shock heating since it is already of approximately the correct temperature when launched.  
However, the association of X-ray emission with shocked material at other wavelengths makes this explanation less likely but would be another possibility explaining the constant appearance.

\section{Summary\label{sect:concl}}
Our new, third epoch $Chandra$ observation clearly shows that the process responsible for the X-ray emission in HH~154 is constant over at least one decade. The position, the luminosity and temperature of the X-ray emission are virtually the same in all observations. Whether differences between the observations are statistical fluctuations or intrinsic differences in the flow cannot be definitely decided due to the low count statistics.
From the trend of the mean energy along the jet axis, we show that the plasma is cooler at larger distances from the driving source.

We discus several models and find that a standing shock most naturally explains the observed morphology given the constant total X-ray luminosity. The location of the X-ray emission, where the outflow is likely collimated, and its stationary appearance argue for this model. Depending on the details of the plasma cooling, the trend in the mean energy can be naturally explained in this model.
The features of the X-ray emission can also be explained in terms of a pulsed jet, where internal shocks cause an apparent stationary X-ray source as the most probable location of an X-ray emitting shock is close to the driving source. The trend in the mean energy might then reflect lower shock velocities or the cooling of the plasma depending on the detailed cooling times of the X-ray emitting plasma. The existence of knots within protostellar jets is usually attributed to time-variable outflows, therefore, such a model is attractive, but it requires a rather regularly modulated flow, since the position, the temperature and the luminosity appear constant.  Variability at larger distances from the driving sources might be present and can be explained either by local shocks or variations of the mass loss rate.

A comparison of our new results for HH~154 with other X-ray emitting jets, in particular with DG~Tau, the only nearby jet X-ray source where multi-epoch observations are available, shows that soft X-ray photons close to the driving source are not unique to HH~154. Therefore, the necessary heating apparently takes place very close to the driving source within the outflow.
With an increasing number of X-ray observations it becomes increasingly clear that the origin of the X-rays is tightly connected to the flow properties within the innermost few 10~AU, either due to inhomogeneities in the outflow or by the collimation process.

New sensitive X-ray observations of HH~154 with a higher cadence are required to decide whether variations on shorter time scales are present and could therefore discriminate between the base-shock and the internal shock model.

\begin{acknowledgements}
      This work has made use of data obtained by $Chandra$, from the $Chandra$ data archive and from the XMM-Newton data archive.
      P.C.S. acknowledges support from the DLR under grant 50OR0703.
      H.M.G. acknowledges support from Chandra under grant GO6-7017X 
      This research has made use of the SIMBAD database, operated at CDS, Strasbourg, France 
\end{acknowledgements}

\bibliographystyle{aa}
\bibliography{cs_bib.bib}
\end{document}